\documentclass[review]{fcs}
\usepackage{layout}
\usepackage{subfig}
\usepackage{hyperref}

\usepackage{forest,adjustbox,xcolor}
\usepackage{pifont}

\usepackage{multirow}   
\usepackage{tabularx}   
\usepackage{booktabs}   

\title{Beyond Self-Talk: A Communication-Centric Survey of LLM-Based Multi-Agent Systems}

\author[1]{Bingyu~YAN}
\author[2]{Zhibo~ZHOU}
\author[4]{Litian~ZHANG}
\author[1]{Lian~ZHANG}
\author[1]{Ziyi~ZHOU}
\author[1]{Dezhuang~MIAO}
\author[3]{Zhoujun~LI}
\author[4]{Chaozhuo~LI}
\author[1,+]{Xiaoming~ZHANG}
\address[1]{School of Cyber Science and Technology, Beihang University, Beijing 100191, China}
\address[2]{The College of Cyber Security/College of Information Science and Technology, Jinan University, Guangzhou, China}
\address[3]{School of Computer Science and Engineering, Beihang University, Beijing 100083, China}
\address[4]{School of Cyber Science and Technology, Beijing University of Posts and Telecommunications, Beijing 100876, China}

\corremail{yolixs@buaa.edu.cn}

\begin{abstract}

Large language model-based multi-agent systems have recently gained significant attention due to their potential for complex, collaborative, and intelligent problem-solving capabilities. Existing surveys typically categorize LLM-based multi-agent systems (LLM-MAS) according to their application domains or architectures, overlooking the central role of communication in coordinating agent behaviors and interactions. To address this gap, this paper presents a comprehensive survey of LLM-MAS from a communication-centric perspective. Specifically, we propose a structured framework that integrates system-level communication (architecture, goals, and protocols) with system internal communication (strategies, paradigms, objects, and content), enabling a detailed exploration of how agents interact, negotiate, and achieve collective intelligence. Through an extensive analysis of recent literature, we identify key components in multiple dimensions and summarize their strengths and limitations. In addition, we highlight current challenges, including communication efficiency, security vulnerabilities, inadequate benchmarking, and scalability issues, and outline promising future research directions. This review aims to help researchers and practitioners gain a clear understanding of the communication mechanisms in LLM-MAS, thereby facilitating the design and deployment of robust, scalable, and secure multi-agent systems.

\end{abstract}

\begin{document}
\section{Introduction}\label{intro}
Large language models (LLMs) have recently demonstrated significant potential across diverse domains. Building upon these strengths, LLMs have been integrated into autonomous agents equipped with profiling, memorization, planning, and action modules~\cite{llm_agent_define}. To make agents more coordinated and scalable when handling complex or dynamic tasks~\cite{single_limit}, LLM-based multi-agent systems (LLM-MAS) have been proposed, where multiple agents interact to achieve goals that exceed the capacity of a single agent. Recent studies underscore LLM-MAS effectiveness in contexts ranging from social simulation\cite{social_media_regulation}, to software engineering\cite{metagpt}, and recommendation systems~\cite{jd_recommendation_system}, illustrating the growing demand for more coordinated and intelligent multi-agent systems (MAS).

\begin{figure}[htbp]
    \centering
    \includegraphics[width=\columnwidth]{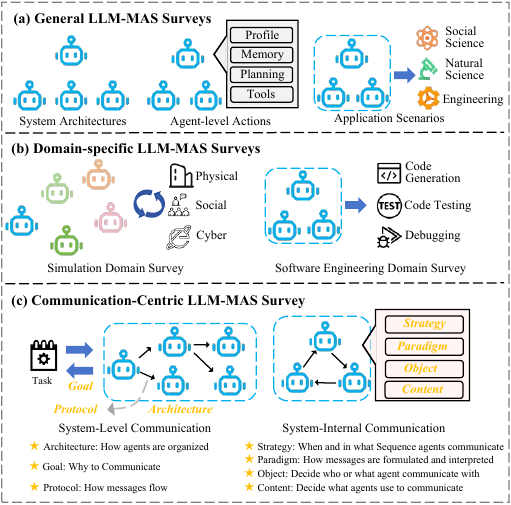} 
    \caption{Position of this survey within existing LLM-MAS literature. The figure contrasts prior general and domain-specific surveys with the communication-centric perspective, underscoring the need for a cross-cutting framework that spans tasks, architectures, workflows, and interaction between agents} 
    \label{fig:introduction} 
\end{figure}

Given the broad application scope and research potential of LLM-MAS, several surveys have appeared to provide researchers with a systematic overview of this emerging research area. These surveys can be broadly categorized into two types: general surveys and domain-specific surveys. As shown in \cref{fig:introduction} (a), general surveys typically focus on system architectures, agent-level actions, and potential application scenarios~\cite{mas_survey_1,mas_survey_2,mas_survey_3}. However, they often provide limited coverage of the communication and coordination workflows among agents, which are critical for effective multi-agent collaboration. As shown in \cref{fig:introduction} (b), domain-specific surveys delve into particular use cases such as social simulation~\cite{mas_application_survey_simulation} or software engineering~\cite{mas_application_survey_1,mas_application_survey_2}. While offering in-depth analyses of domain-specific LLM-MAS workflows, these studies lack generalizability and fail to propose a comprehensive framework for broader applications.


Inter-agent communication plays a critical role in enabling LLM-MAS to perform more complex tasks compared to single-agent systems by facilitating idea exchange and coordinated planning. Drawing on insights from traditional communication theory~\cite{communication_1,communication_2}, concepts such as "source" and "channel" align closely with the communication processes present in the LLM-MAS workflow. Consequently, as shown in \cref{fig:introduction} (c), the LLM-MAS workflow can be effectively decomposed through a communication-centric perspective. Specifically, we define LLM-MAS as a communication protocol-constrained automated system driven by communication goals within a predefined communication architecture. Agents in this system have multiple communication strategies and paradigms, interacting with various communication objects to exchange diverse content for task completion.


\cref{fig:llm-mas-taxonomy} summarizes the structure of this survey. To provide necessary context, Section~\ref{sec:background} introduces foundational concepts and terminology relevant to LLM-MAS. Building upon this, we propose a two-level analytical framework distinguishing between system-level communication and system-internal communication, discussed respectively in Section~\ref{sec:system-level} and Section~\ref{sec:system-internal}. The system-level communication analysis addresses how agents are organized, their overarching communication goals, and key emerging protocols. In contrast, the system-internal communication analysis delves deeper into the internal communication dynamics among agents, examining their strategies, paradigms, communication objects, and exchanged content. Subsequently, Section~\ref{sec:challenges} highlights current challenges and outlines promising research opportunities. This structure provides a holistic understanding of the workflow across varied tasks.

\begin{figure*}[ht]
  \centering
  \begin{adjustbox}{width=2\columnwidth}
    \begin{forest}
      for tree={
        rounded corners,
        child anchor=west,
        parent anchor=east,
        grow'=east,
        draw=blue!60!black,
        anchor=west,
        node options={align=center,font=\small},
        edge path={
          \noexpand\path[\forestoption{edge}]
            (.child anchor) -| +(-5pt,0) -- +(-5pt,0) |-
            (!u.parent anchor)\forestoption{edge label};
        },
        where level=0{text width=0.8cm}{},
        where level=1{text width=3.8cm}{},
        where level=2{text width=4cm}{},
        where level=3{text width=7cm,draw=blue!40!black,fill=blue!10}{}
      }
      [\rotatebox{90}{Communication-Centric LLM-MAS}, align=center
        [Background \S~\ref{sec:background}
          [Single-Agent Systems \S\ref{subsec:single} ,text width=6cm]
          [Traditional Multi-Agent Systems  \S\ref{subsec:traditional_mas},text width=6cm]
          [LLM-Based Multi-Agent Systems  \S\ref{subsec:llm_mas},text width=6cm]
        ]
        [System-Level \\Communication \S\ref{sec:system-level}
          [Communication Architecture \S\ref{subsec:architecture}
            [{Flat~\ref{para:arch_flat}, Hierarchical~\ref{para:arch_hier}, Team~\ref{para:arch_team}, Society~\ref{para:arch_society}, Hybrid~\ref{para:arch_hybrid}}]
          ]
          [Communication Goal \S\ref{subsec:goal}
            [{Cooperation~\ref{para:goal_cooperation}, Competition~\ref{para:goal_competition}, Mixed~\ref{para:goal_mix}}]
          ]
          [Communication Protocol \S\ref{subsec:protocol}
            [{MCP~\ref{para:MCP}, A2A~\ref{para:A2A}, ANP~\ref{para:ANP}}]
          ]
        ]
        [System Internal \\Communication \S\ref{sec:system-internal}
          [Communication Strategy \S\ref{subsec:strategie}
            [{One-by-One~\ref{para:stra_onebyone}, Simultaneous-Talk~\ref{para:stra_aimu}, \\Simultaneous-Talk with Summarizer~\ref{para:stra_simu_summarizer}}]
          ]
          [Communication Paradigm \S\ref{subsec:paradigm}
            [{Message Passing~\ref{para:paradigm_messagepassing}, Speech Act~\ref{para:paradigm_speechact}, Blackboard~\ref{para:paradigm_blackboard}}]
          ]
          [Communication Object \S\ref{subsec:object}
            [{Self~\ref{para:object_self}, Other Agents~\ref{para:object_agent}, \\Environment~\ref{para:object_environment}, Human~\ref{para:object_human}}]
          ]
          [Communication Content \S\ref{subsec:content}
            [{Explicit (Natural Language, Code \& Data)~\ref{para:content_explicit},\\
            Implicit (Behavioral Feedback, Environment Signal)~\ref{para:content_implicit}}]
          ]
        ]
        [Challenges \& Opportunities \S\ref{sec:challenges}
          [Optimizing the System Design \S\ref{subsec:opt},text width=6cm]
          [Advancing Research on Agent Competition \S\ref{subsec:competition},text width=6cm]
          [Unified Communication Protocol \S\ref{subsec:unified_protocol},text width=6cm]
          [Multimodal Communication \S\ref{subsec:multimodal},text width=6cm]
          [Communication Security \S\ref{subsec:security},text width=6cm]
          [Benchmarks \& Evaluation \S\ref{subsec:benchmark},text width=6cm]
        ]
      ]
    \end{forest}
  \end{adjustbox}
  \caption{The structure of this paper}
  \label{fig:llm-mas-taxonomy}
\end{figure*}

Our main contributions include:
\begin{itemize}
    \item \textbf{Comprehensive Framework:} From a communication perspective, we designed a comprehensive framework applicable to all types of LLM-MAS, analyzing the workflow at both the system level and the internal system level.
    \item \textbf{Deep Analysis of Communication Processes:} We dissect real-world examples and prototypes to illustrate how well-orchestrated communication leads to more effective multi-agent behavior.
    \item \textbf{Identification of Challenges and Opportunities:} We shed light on open issues like scalability, security, and multimodal integration, and offer potential research directions for both academia and industry.
\end{itemize}
\section{Background}\label{sec:background}
This chapter provides the conceptual grounding for the remainder of the survey. We first revisit the internal structure and operating principles of LLM‑based single agents. Then introducing the traditional multi‑agent systems (MAS). We conclude by articulating how LLM‑MAS inherit, extend, and occasionally challenge the assumptions of earlier paradigms, thereby motivating the communication-centric taxonomy developed in subsequent sections.

\subsection{LLM-Based Agents}\label{subsec:single}
LLM‑based agents are autonomous entities whose core reasoning component is a LLM. They represent the cornerstone of multi-agent architectures, as they lay the groundwork for understanding how individual agents reason, plan, and act. In line with the taxonomy proposed by ~\cite{agent_survey_1}, this section outlines the composition and functionalities of LLM-based agents.

An LLM-based agent typically consists of three key components: \textit{Brain}, \textit{Perception}, and \textit{Action}, each playing a distinct yet complementary role. 1) \textbf{Brain} is a LLM that integrates both short-term and long-term memory modules to reduce potential hallucinations and enhance the agent's reasoning and planning capabilities. Short-term memory caches the current conversational context while long-term memory stores structured artefacts accumulated across sessions. Various retrieval augmented generation (RAG) methods~\cite{rag_1} and RAG enhancement research~\cite{rag_2} play an important role in enhancing memory modules. 2) \textbf{Perception} is a multimodal interface that converts raw sensory inputs including text, vision, audio, or environmental state vectors into prompts consumable by the Brain. 3) \textbf{Actions} are not limited to generating textual responses, they also encompass integration with external tools such as web APIs or real-world actuators~\cite{toolllm,embodied_agents} to enhance the agent’s problem-solving capabilities.

\subsection{Traditional Multi-Agent Systems}\label{subsec:traditional_mas}
Before the rise of LLMs, the MAS research had established a rich toolbox spanning symbolic reasoning and learning-based coordination \cite{traditionalmas}. Early symbolic MAS describe agents in terms of Beliefs, Desires, and Intentions and prescribe communication via agent-communication languages. These languages provide machine-interpretable speech acts such as inform, request and propose, and support standard protocol templates including contract nets and auction-based negotiation, thereby ensuring transparent reasoning about commitments, goals, and obligations. However, their discrete vocabularies and brittle parsers make them ill-suited for open-domain tasks.

The learning-based MAS replaces handcrafted rules with optimisation. Centralised-training-with-decentralised-execution frameworks~\cite{mas_ctde} use a global critic to stabilise gradients while permitting fully distributed policies at test time. Independent Q-learning and policy-gradient variants alleviate non-stationarity, whereas value-decomposition, counterfactual baselines, and influence-based credit assignment attack the credit-assignment problem directly. Graph neural agents, in turn, convey compact latent messages over dynamic communication graphs and scale to hundreds of cooperative robots, traffic-signal controllers, or swarm drones.  Nonetheless, most learning-based systems still adopt narrow, task-specific message formats, incur heavy sample complexity, and provide limited interpretability which motivated the search for language-native coordination.

\subsection{LLM-Based Multi-Agent Systems}\label{subsec:llm_mas}
While single-agent systems exhibit strong individual reasoning, they often struggle with tasks requiring collective intelligence or large-scale coordination. MAS~\cite{mas_define} can address these limits by orchestrating multiple intelligent agents and leveraging communication as a key mechanism for goal alignment. Meanwhile, LLM-MAS combine the linguistic flexibility of LLM agents with the coordination principles of MAS, thereby dissolving the boundary between natural-language dialogue and agent communication. Agents can cooperate, compete, or negotiate, depending on system objectives and architectural choices. Such flexibility in communication design sets the stage for LLM-MAS, where advanced language models enable more sophisticated inter-agent interactions across diverse application domains.

\begin{table*}[htbp]
  \centering
  \small               
  \caption{Representative LLM-MAS studies grouped by communication architecture and communication goal}
  \begin{tabularx}{\textwidth}{c c c}
    \toprule
    \textbf{Architecture} & \textbf{Goal} & \textbf{Work} \\
    \midrule

    \multirow{3}{*}{Flat}
      & Cooperation &
        Ulmer et al.~\cite{boostrapping};
        Abdullin et al.~\cite{dataset_generation};
        Du et al.~\cite{debate_improve_llm_1};
        Group Think~\cite{groupthink}\\

      & Competition &
        GAMA\textendash Bench~\cite{evaluating_llm_game_ability} \\

      & Mixed &
        Chuang et al.~\cite{simulating_opinion_dynamic};
        GameChat~\cite{gamechat}\\

    \midrule
    \multirow{5}{*}{Hierarchical}
      & \multirow{3}{*}{Cooperation} &
        ChatDev~\cite{chatdev_software_development};
        AutoDefense~\cite{autodefense_against_jailbreak};
        CausalGPT~\cite{casualgpt_reasoning};
        SoA~\cite{soa_code_generation};
        FINCON~\cite{fincon_decision_making};
        Wang et al.~\cite{mas_benchmark};
        CoA~\cite{chain_of_agents};\\
      &  &
        AutoData~\cite{autodata};
        AutoHMA-LLM~\cite{autohma};
        G-Safeguard~\cite{gsafeguard};
        Li et al.~\cite{li2025knowledge};
        Kalyuzhnaya et al.~\cite{kalyuzhnaya2025llm};
        LogiAgent~\cite{logiagent};\\
      &  &
        MACM~\cite{macm};
        CPDESIM~\cite{codesim};
        Shen et al.~\cite{shen2025optimizing};
        EvoMAC~\cite{evomac};
        Wang et al.~\cite{wang2025talk}\\

      & Competition &
        Cai et al.~\cite{social_media_regulation} \\

      & Mixed &
        Ju et al.~\cite{community_knowledge_flooding};
        Wu et al.~\cite{mas_for_jvbensha} \\

    \midrule
    \multirow{3}{*}{Team}
      & \multirow{2}{*}{Cooperation} &
        MetaGPT~\cite{metagpt};
        MAGIS~\cite{magis_mas_for_github};
        AgentCoord~\cite{agentcoord};
        AutoAgents~\cite{autoagents};
        Qian et al.~\cite{mas_for_software_2};
        SimClass~\cite{classroom_simulation};
        Fan et al.~\cite{fan2025llm};\\
      &  & 
        Ma et al.~\cite{ma2025communication};
        FactGuard~\cite{factguard};
        Luo et al.~\cite{luo2025llm};
        MDTeamGPT~\cite{mdteamgpt};
        MedAgents~\cite{medagents};
        Thematic-LM~\cite{thematiclM}\\

      & Mixed &
        POLCA~\cite{polca_mas_for_political};
        ReCon~\cite{recon_thinking} \\

    \midrule
    \multirow{2}{*}{Society}
      & Cooperation &
        Generative Agents~\cite{stanf_villege};
        GOVSIM~\cite{govsim};
        Yue et al.~\cite{shengbinyue2025multi};
        Sreedhar et al.~\cite{sreedhar2025simulating}\\

      & Mixed &
        Dai et al.~\cite{sct_society};
        EconAgent~\cite{econagent};
        Guan et al.~\cite{richeliey_diplomacy_society};
        EcoLANG~\cite{ecolang}\\

    \midrule
    \multirow{4}{*}{Hybrid}
      & \multirow{2}{*}{Cooperation} &
        FixAgent~\cite{fixagent_mas_for_debug};
        PeerGPT~\cite{peergpt};
        ChatSim~\cite{chatsim_mas_make_scene};
        Tulip Agent~\cite{mas_use_tool};
        MADR~\cite{debate_2};
        AgentFM~\cite{agentfm};\\
     &  &
        Ronanki et al.~\cite{ronanki2025facilitating};
        Zhang et al.~\cite{zhang2025facilitating};
        Flow-of-Action~\cite{flow-of-action};
        Wang et al.~\cite{wang2025llm}\\

      & Competition &
        Agent4Debate~\cite{agent4debate};
        Zhang et al.~\cite{mas_for_poetry_generation} \\

      & Mixed &
        Wu et al.~\cite{shallwetalk};
        BlockAgents~\cite{blockagents};
        Xu et al.~\cite{mas_for_defence_attack};
        ToM~\cite{mas_for_guandan};
        TED~\cite{TED}\\

    \bottomrule
  \end{tabularx}

  \label{tab:applications-llm-mas}
\end{table*}

\section{System-Level Communication}\label{sec:system-level}
In this section, the LLM-MAS is analyzed from the macro system-level communication. We first examine the communication architecture, highlighting how different architectures such as flat, hierarchical or society-based organize agents. Each of these architectures offers distinct approaches to agent interaction, with implications for scalability, flexibility, and efficiency in system performance. We then delve into the communication goals that decide why agents communicate, where cooperation, competition, or a mixture of both can lead to distinct multi-agent dynamics. Through specific application examples, we demonstrate the advantages, limitations, and applicable scenarios of these communication architectures and goals. Finally, this section discusses the emerging communication protocols and introduces three prominent protocols in detail to clarify the standardized communication behavior. As shown in Table~\ref{tab:applications-llm-mas}, because the communication protocol has not been widely studied and used, we categorize current LLM-MAS studies according to the communication architectures and communication goals in detail.

\subsection{Communication Architecture}\label{subsec:architecture}
\begin{figure}[htbp]
    \centering
    \includegraphics[width=\columnwidth]{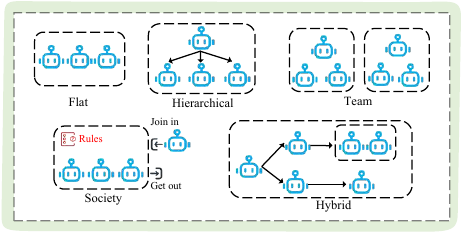} 
    \caption{Five canonical communication architectures for LLM-MAS} 
    \label{fig:architecture} 
\end{figure}

Communication architecture defines the structural organization of agents within a LLM-based multi-agent system and governs the flow of information among them~\cite{mas_a_survey}. It significantly influences both macro-level system behaviors and micro-level interactions, shaping aspects like scalability, flexibility, and operational efficiency. The choice of architecture usually depends on task complexity, agent autonomy levels, and desired interaction patterns. As shown in \cref{fig:architecture}, we explore five primary communication architectures: Flat, Hierarchical, Team, Society, and Hybrid architectures, providing additional detailed discussions and examples to deepen understanding.

\subsubsection{Flat Architecture}\label{para:arch_flat} 

Flat architecture is characterized by a decentralized network of peer-to-peer agent interactions without hierarchical distinctions or central oversight. This architecture excels in environments requiring agile, flexible, and spontaneous interactions among agents, such as dynamic task assignments and rapid decision-making scenarios. For example, Abdullin et al. \cite{dataset_generation} demonstrate flat architecture effectiveness in generating synthetic dialogues, where agents collaboratively and iteratively refine outputs. Similarly, Du et al. \cite{debate_improve_llm_1} employ a flat architecture for fact-checking, enabling agents to critique each other’s reasoning dynamically, thereby enhancing overall accuracy and reducing hallucinations. Despite its benefits in adaptability and responsiveness, a flat structure can encounter challenges in scalability, as increased agent count may lead to increased communication overhead and reduced coordination efficiency.

\subsubsection{Hierarchical Architecture}\label{para:arch_hier}

In hierarchical architectures, agents are structured into layers or tiers, with higher-level agents responsible for strategic oversight and lower-level agents performing detailed execution tasks. This configuration is particularly effective for complex scenarios demanding clear task delegation, supervision, and integration of diverse agent competencies. For instance, ChatDev ~\cite{chatdev_software_development} organizes agents into specialized roles such as designers, coders, and testers under senior management agents, effectively facilitating software development through structured workflows. CausalGPT ~\cite{casualgpt_reasoning} exemplifies hierarchical use by employing higher-level evaluators to guide lower-level agent reasoning, thereby significantly improving output consistency. However, hierarchical systems may encounter bottlenecks, particularly when higher-level agents become overloaded or communication delays between layers emerge.

\subsubsection{Team Architecture.}\label{para:arch_team} 

Team-based architectures segment agents into specialized groups or teams, each focusing on distinct tasks or domains, thus leveraging specific agent expertise and fostering robust intra-team collaboration. For instance, MAGIS ~\cite{magis_mas_for_github} demonstrates the benefits of team architecture by categorizing agents into administrators, developers, and testers, streamlining the collaborative resolution of software issues on platforms such as GitHub. The POLCA framework ~\cite{polca_mas_for_political} further illustrates team architecture versatility by employing specialized teams to simulate political negotiations, predicting coalition formation outcomes. While team architectures effectively leverage specialized expertise, the increased inter-team communication demands can introduce additional overhead, necessitating efficient cross-team coordination mechanisms.

\subsubsection{Society Architecture}\label{para:arch_society} 

Society architectures model agent interactions within broader social environments governed by shared norms or rules, reflecting emergent behaviors characteristic of human societies. These architectures support complex, large-scale simulations involving diverse agents with varying roles, motivations, and interactions. Generative Agents ~\cite{stanf_villege}, for example, create sandbox environments where agents dynamically form friendships, organize activities, and adapt behaviors based on past experiences. EconAgent ~\cite{econagent} integrates economic dynamics into agent behaviors, simulating realistic economic decisions like consumption, investment, and employment, thus providing insights into macroeconomic phenomena driven by micro-level actions. Although society architectures excel at simulating emergent phenomena and social interactions, they can become challenging to manage due to the inherent complexity and unpredictability of agent behaviors.

\subsubsection{Hybrid Architecture}\label{para:arch_hybrid} 

Not all designs fit neatly into a particular architecture, and some architectures may contain features from several different styles. Hybrid architectures combine multiple communication structures to create flexible and adaptable multi-agent systems capable of addressing diverse challenges. By integrating hierarchical oversight with decentralized agent collaboration, hybrid architectures optimize agent interactions and resource allocation dynamically. FixAgent ~\cite{fixagent_mas_for_debug} exemplifies hybrid architectures by adapting hierarchical structures to debugging tasks, combining strategic oversight with decentralized task execution. ChatSim ~\cite{chatsim_mas_make_scene} integrates top-down control for scene rendering with peer-to-peer collaboration among agents for detailed asset integration, thus overcoming limitations inherent to purely hierarchical or flat systems. Tulip Agent ~\cite{mas_use_tool} employs a hybrid approach enabling agents to dynamically discover and utilize tools, showcasing enhanced flexibility, adaptability, and scalability across multiple scenarios. However, hybrid systems require sophisticated design strategies to avoid coordination overhead and ensure efficient operation.

In summary, the choice of communication architecture significantly influences the capabilities, efficiency, and scalability of LLM-based multi-agent systems. Researchers and practitioners must carefully evaluate their specific requirements, considering trade-offs between flexibility, control, scalability, and efficiency, to select the most suitable architecture or combination thereof.

\subsection{Communication Goal}\label{subsec:goal}
\begin{figure}[htbp]
    \centering
    \includegraphics[width=\columnwidth]{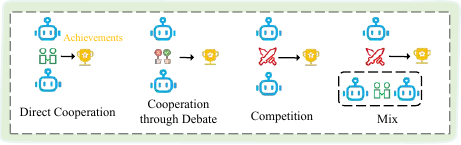} 
    \caption{Communication goals driving multi-agent interaction} 
    \label{fig:goal} 
\end{figure}

Communication goals represent the intended purpose behind agent interactions, driving behaviors, information-sharing patterns, and coordination strategies within a multi-agent system. Clearly articulating these goals enhances system efficiency, coherence, and adaptability to diverse scenarios, as they encapsulate the “why” behind the communication. As shown in \cref{fig:goal}, this section elaborates on three primary communication goals: Cooperation (including Direct Cooperation and Cooperation through Debate), Competition, and Mixed goals.

\subsubsection{Cooperation}\label{para:goal_cooperation} 
Cooperation is a central feature in numerous multi-agent scenarios, which involves agents collaboratively working towards shared objectives, leveraging collective knowledge and skills to achieve outcomes superior to those achievable individually. Within cooperative settings, interactions are often characterized by transparency, mutual trust, and a focus on common goals. Cooperation is divided into two main types:
direct cooperation and cooperation through debate.

Direct cooperation refers to straightforward and seamless collaboration where agents openly share information and resources to jointly complete tasks or solve problems. This interaction style emphasizes rapid consensus and efficient integration of expertise, critical in time-sensitive or complex problem-solving environments. For instance,   some works harness direct cooperation to enhance LLM reasoning\cite{boostrapping,casualgpt_reasoning,autoagents},  while others employ it in collaborative code generation \cite{chatdev_software_development,soa_code_generation,magis_mas_for_github,mas_for_software_2,fixagent_mas_for_debug} and scientific experimentation\cite{chatsim_mas_make_scene,mas_use_tool}. The research community continues to explore advanced coordination models~\cite{agentcoord,govsim,mas_benchmark} for fine-grained task allocation, robust communication protocols, and scalability in agent collectives.

Cooperation through debate introduces a nuanced cooperative model where agents actively critique and refine each other's inputs through structured debates. Rather than passively accepting shared information, agents rigorously challenge each other's assumptions, reasoning, and outcomes to achieve more robust and accurate conclusions. Examples include factual reasoning\cite{debate_improve_llm_1} and fact-checking\cite{debate_2}, where rigorous back-and-forth discussions help validate information and reduce the likelihood of mistakes. This method effectively balances collaborative interaction with critical evaluation, producing more resilient and well-rounded outcomes.

\subsubsection{Competition}\label{para:goal_competition} Competitive interactions in LLM-based multi-agent systems occur when agents have conflicting objectives or vie for limited resources or favorable outcomes. This adversarial context stimulates strategic thinking, innovation, and adaptive behaviors, as agents must continually refine their strategies to succeed. 

The game-playing ability of agents in the system has been demonstrated in research GAMA–Bench~\cite{evaluating_llm_game_ability}. In these adversarial contexts, agents actively shape the interaction by generating persuasive or deceptive narratives. For instance, the study by Cai et al.~\cite{social_media_regulation} explores language evolution within adversarial simulations, providing insights into how competitive interactions drive linguistic adaptation. Agent4Debate~\cite{agent4debate} designed a dynamic multi-agent framework for competitive debate with humans. One significant advantage of competition in LLM-MAS is its potential to drive innovation and strategic sophistication. As agents continually adjust to rivals’ tactics and environmental feedback, they refine their own approaches in a never-ending loop of adaptation. For instance,~\cite{mas_for_poetry_generation} explored a poetry-generation task where agents strove to outperform each other through creative linguistic outputs.

However, competitive settings introduce a range of challenges, including potential instability and security vulnerabilities induced by adversarial or deceptive agent behaviors. Therefore, it is imperative to design mechanisms that carefully reconcile competitive incentives with cooperative regulatory constraints to ensure sustainable and robust system performance.

\subsubsection{Mixed}\label{para:goal_mix} 

While purely cooperative or purely competitive settings offer conceptual clarity, mixed communication goals blend cooperative and competitive elements, reflecting the complexity and fluidity of real-world interactions. Within these mixed regimes, agents continuously reconstitute alliances, vie for scarce resources, and initiate strategic collaborations as situational contingencies and task objectives evolve. Such flexibility enhances the system’s capacity to tackle multi-faceted problems and to maintain robust performance under dynamically changing environmental conditions.

In practice, such mixed interactions are pivotal for capturing nuanced, dynamic behavior in domains like social simulations~\cite{community_knowledge_flooding,sct_society}, collective problem-solving, or negotiation tasks~\cite{polca_mas_for_political,richeliey_diplomacy_society}. Allowing agents to strike a balance between cooperative and adversarial behavior makes it possible to emerge solutions that cannot be suggested by purely cooperative or competitive environments.

Despite their versatility, mixed-goal interactions introduce complexities regarding mechanism design, protocol standardization, and ethical considerations. Balancing cooperation and competition dynamically necessitates sophisticated coordination and monitoring mechanisms to ensure system stability and ethical integrity.

In summary, clearly defining and structuring communication goals is critical to optimizing multi-agent interactions. Selecting and implementing appropriate goals depend heavily on task requirements, system scalability, and desired agent dynamics, ultimately influencing overall system effectiveness.

\subsection{Communication Protocol}\label{subsec:protocol}
\begin{figure}[htbp]
    \centering
    \includegraphics[width=\columnwidth]{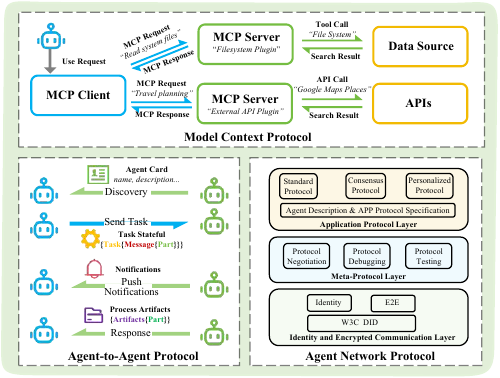} 
    \caption{Emerging communication protocols for LLM-MAS} 
    \label{fig:protocol} 
\end{figure}

While the communication architecture decides who can talk to whom and the communication goal clarifies why they talk, a dedicated communication protocol specifies how the messages actually flow between agents and external systems. Standardized protocols are now emerging that make LLM-MAS deployments portable, secure, and easier to audit. Communication protocols detail the actual mechanics of interaction, ensuring consistent, secure, and efficient communication. As shown in \cref{fig:protocol}. This section introduces three key emerging protocols: Model Context Protocol (MCP)~\cite{mcp}, Agent-to-Agent Protocol (A2A)~\cite{a2a}, and Agent Network Protocol (ANP)~\cite{anp}.

\subsubsection{Model Context Protocol}\label{para:MCP}
MCP is a general-purpose context-oriented protocol developed to facilitate secure and structured interactions between LLM agents and external resources such as tools, data, and services. MCP utilizes a JSON-RPC client-server architecture, enabling agents (hosts) to request and receive context from external resources through intermediate components—clients and servers. In this architecture, the host initiates context requests based on user queries, and the client manages connections to both the host and server, providing resource descriptions and executing context requests. The server directly interacts with resources to execute client requests and relay context, while resources include data, tools, or services. This layered approach reduces fragmentation across various agents and tool providers, significantly improving interoperability and scalability. Additionally, MCP enhances data security by decoupling sensitive tool invocations from LLM-generated responses, thus minimizing the risks of data exposure.

\subsubsection{Agent-to-Agent Protocol}\label{para:A2A}
A2A supports inter-agent communication, emphasizing secure and structured peer-to-peer interactions among LLM agents. A2A employs capability-based "Agent Cards" distributed via HTTP and Server-Sent Events, facilitating efficient task outsourcing and collaboration within enterprise-scale deployments. Agents advertise their capabilities through Agent Cards, enabling dynamic task delegation based on real-time capability awareness. Communication occurs directly between peers without centralized intermediaries, significantly reducing latency and improving responsiveness. Additionally, asynchronous messaging via server-sent events enhances scalability and robustness. A2A's structured, capability-driven interactions improve task efficiency and collaboration fidelity, particularly suited for complex enterprise environments.

\subsubsection{Agent Network Protocol}\label{para:ANP}
ANP facilitates decentralized agent communication and discovery over open networks. Built upon decentralized identifiers (DIDs) and JSON-LD graphs, ANP promotes secure, interoperable interactions among heterogeneous agents. Agents dynamically discover and verify peers using DIDs, ensuring secure identification across network boundaries. Communication is structured via JSON-LD graphs, enabling semantic clarity and context-aware interactions. ANP also incorporates encryption and secure channels, safeguarding data integrity and confidentiality. Its decentralized infrastructure supports robust, cross-organizational collaboration, fostering scalable and secure multi-agent ecosystems on a global scale.

Table \ref{tab:protocol} adapts the comparative framework proposed by Abul et al.~\cite{protocolsurvey}, complements it and extends it with additional dimensions such as security and use cases to better contextualise LLM-MAS. MCP, A2A, and ANP represent significant advancements in standardizing and enhancing communication within LLM-based multi-agent systems. Additionally, other emerging protocols such as Agent Communication Protocol (ACP)~\cite{acp}, Agent Interaction and Task Protocol (AITP)~\cite{aitp}, and Agent Content Protocol (AConP)~\cite{aconp} are also gaining attention, highlighting the dynamic and evolving landscape of communication protocols. Collectively, these protocols address interoperability, security, scalability, and context management, paving the way toward more integrated and intelligent multi-agent deployments.

~
\newcolumntype{A}{>{\raggedright\arraybackslash}X} 
\setlength{\tabcolsep}{2pt}
\begin{table}[h]
  \centering
  \footnotesize
  \caption{Comparison of Three Emerging Agent Protocols}
  \label{tab:protocol}
  \begin{tabularx}{\linewidth}{l *{3}{A}}
    \toprule
    \textbf{Aspect} & \textbf{MCP} & \textbf{A2A} & \textbf{ANP} \\
    \midrule
    Topology &
    JSON\=/RPC over client--server channel &
    Peer-style client $\leftrightarrow$ remote agent link &
    Decentralised P2P agent mesh \\
    \addlinespace
    Discovery &
    Static endpoint or manual registry entry &
    Signed \emph{Agent Card} retrieval via HTTP &
    Search-index crawling + DID document exchange \\
    \addlinespace
    Format &
    Typed JSON-RPC 2.0 messages &
    Task \& Artifact bundles &
    JSON-LD; meta-protocol negotiation \\
    \addlinespace
    Security &
    Bearer tokens; optional DID claims; RBAC &
    OAuth 2 / enterprise IAM; mutual TLS &
    DID-based handshake; end-to-end encryption \\
    \addlinespace
    Strengths &
    Seamless LLM tool calling; minimal bootstrap &
    Rich task delegation; vendor-backed ecosystem &
    Trustless identity; no single point of control \\
    \addlinespace
    Limitations &
    Centralised service dependency; safety risks &
    Cross-team workflows; delegated multi-step automation &
    P2P logistics coordination; DAO-driven negotiation \\
    \addlinespace
    Use Cases &
    Plugin APIs; retrieval-augmented Q\&A &
    Enterprise workflows; delegated multi-step tasks &
    P2P logistics; DAO/DeFi negotiation \\
    \bottomrule
  \end{tabularx}
\end{table}

\section{System Internal Communication}\label{sec:system-internal}
Having established the overall system-level perspective, we now turn to the internal communication within the LLM-MAS. We first categorize and compare diverse communication strategies that determine when and in what sequence agents communicate. Next, we discuss the communication paradigms that frame the form and modality of these exchanges, followed by an examination of communication objects and content. These elements collectively define the richness, adaptability, and effectiveness of communication in LLM-MAS, ultimately shaping the system’s capacity for coordination and problem-solving.

\newcolumntype{B}{>{\centering\arraybackslash}X}

\begin{table*}[htbp]
  \centering
  \footnotesize
  \caption{Summary of current LLM-MAS studies with their \textit{Communication Strategy}, \textit{Communication Paradigm}, \textit{Communication Object}, and \textit{Communication Content}.  
  For the strategy module, OO, ST, SS denote One-by-one, Simultaneous-Talk, Simultaneous-Talk-with-Summarizer; For the paradigm module, 
  MP, SA, BB denote message passing, speech act, blackboard; For the object module, SE, OA, EN, HU denote self, other agents, environment, human; For the content module, EC, IC denote explicit and implicit content}
  \label{tab:addlabel_trimmed}

    \begin{tabularx}{\textwidth}{l *{5}{B}}
      \hline
      \textbf{Work} & \textbf{Strategy} & \textbf{Paradigm} & \textbf{Object} & \textbf{Content} & \textbf{Time} \\
      \hline   
      Generative Agents~\cite{stanf_villege}      & ST, SS & MP    & OA, EN & EC, IC & 04/2023\\
      Du et al.~\cite{debate_improve_llm_1}      & ST          & MP, SA & OA             & EC, IC & 05/2023\\
      ChatDev~\cite{chatdev_software_development} & OO          & MP, SA & OA             & EC, IC & 07/2023\\
      MetaGPT~\cite{metagpt}                      & SS          & BB             & OA, EN & EC & 08/2023\\
      CausalGPT~\cite{casualgpt_reasoning}         & OO, SS & MP, BB & OA, EN           & EC & 08/2023\\
      AutoAgents~\cite{autoagents}                & ST          & MP             & OA             & EC & 09/2023\\
      EconAgent~\cite{econagent}                  & ST          & MP             & OA, EN & EC, IC & 10/2023\\
      Chuang et al.~\cite{simulating_opinion_dynamic} & ST       & MP. SA & OA             & EC, IC & 11/2023\\
      MedAgents~\cite{medagents}                  & OO, ST          & MP             & SE, OA, HU & EC, IC & 11/2023\\
      Wu et al.~\cite{mas_for_jvbensha}           & OO          & MP             & SE, OA, EN & EC, IC & 12/2023\\
      Qian et al.~\cite{mas_for_software_2}       & OO          & MP, BB & OA             & EC & 12/2023\\
      Ulmer et al.~\cite{boostrapping}           & OO           & MP             & SE, OA &EC & 01/2024\\
      Abdullin et al.~\cite{dataset_generation}  & OO           & MP             & OA             & EC & 01/2024\\
      Dai et al.~\cite{sct_society}               & ST          & MP             & OA             & EC, IC & 01/2024\\
      Wang et al.~\cite{mas_benchmark}            & OO         & MP             & OA             & EC, IC & 02/2024\\
      POLCA~\cite{polca_mas_for_political}        & SS          & MP, SA & OA             & EC & 02/2024\\
      ChatSim~\cite{chatsim_mas_make_scene}       & SS          & MP, SA & OA             & EC & 02/2024\\
      MADR~\cite{debate_2}                        & SS          & MP             & OA             & EC & 02/2024\\
      Wu et al.~\cite{shallwetalk}                & OO, ST & MP          & OA             & EC, IC & 02/2024\\   
      GAMA-Bench~\cite{evaluating_llm_game_ability} & ST        & MP             & OA             & EC & 03/2024\\
      PeerGPT~\cite{peergpt}                      & OO, SS & SA          & OA, HU & EC & 03/2024\\  
      AutoDefense~\cite{autodefense_against_jailbreak} & OO, ST & MP         & OA            & EC & 03/2024\\
      AgentCoord~\cite{agentcoord}                & SS          & MP, SA & SE, OA, EN, HU & EC, IC & 04/2024\\
      MACM~\cite{macm}                  & OO          & MP             & SE, OA, EN & EC & 04/2024\\
      SoA~\cite{soa_code_generation}              & OO          & MP            & OA             & EC & 04/2024\\
      Xu et al.~\cite{mas_for_defence_attack}     & ST          & MP             & OA            & EC & 04/2024\\
      GOVSIM~\cite{govsim}                        & SS          & MP             & OA             & EC & 04/2024\\
      FixAgent~\cite{fixagent_mas_for_debug}      & SS          & MP            & OA            & EC & 04/2024\\
      MAGIS~\cite{magis_mas_for_github}           & OO          & MP            & OA, EN & EC & 05/2024\\
      Cai et al.~\cite{social_media_regulation}   & OO, SS & MP, SA & SE, OA, EN & EC, IC & 05/2024\\
      CoA~\cite{chain_of_agents}                  & OO          & MP             & OA             & EC & 06/2024\\
      FINCON~\cite{fincon_decision_making}        & SS          & MP             & OA, EN & EC & 07/2024\\
      SimClass~\cite{classroom_simulation}        & SS          & MP            & OA            & EC & 07/2024\\
      BlockAgents~\cite{blockagents}              & SS          & MP, SA & OA             & EC, IC & 07/2024\\
      Tulip Agent~\cite{mas_use_tool}            & SS          & MP            & OA            & EC, IC & 07/2024\\
      Guan et al.~\cite{richeliey_diplomacy_society} & ST        & MP, SA & OA            & EC, IC & 07/2024\\ 
      Ju et al.~\cite{community_knowledge_flooding}& OO         & MP             & OA             & EC, IC & 07/2024\\
      Agent4Debate~\cite{agent4debate}            & SS         & MP             & OA, EN & EC & 08/2024\\
      ToM~\cite{mas_for_guandan}                  & ST          & MP             & OA, EN & EC & 08/2024\\
      Zhang et al.~\cite{mas_for_poetry_generation} & ST         & MP             & OA             & EC & 09/2024\\
      Li et al.~\cite{li2025knowledge}                  & SS          & MP             & OA, EN & EC & 09/2024\\
      Kalyuzhnaya et al.~\cite{kalyuzhnaya2025llm}                  & SS         & MP            & OA, EN, HU & EC & 10/2024\\
      EvoMAC~\cite{evomac}                  & SS          & MP, SA             & OA, EN & EC, IC & 10/2024\\
      ReCon~\cite{recon_thinking}                 & SS         & MP, SA & OA             & EC, IC & 11/2024\\ 
      Fan et al.~\cite{fan2025llm}                  & ST         & MP, SA             & OA, EN & EC & 01/2025\\
      Wang et al.~\cite{wang2025llm}                  & SS          & MP             & SE, OA, HU & EC & 01/2025\\
      Luo et al.~\cite{luo2025llm}                  & SS         & MP            & SE, OA, EN & EC & 01/2025\\
      Ma et al.~\cite{ma2025communication}                  & OO          & MP, SA             & OA & EC & 02/2025\\
      Sreedhar et al.~\cite{sreedhar2025simulating}                  & OO          & MP             & OA, EN & EC & 02/2025\\
      G-Safeguard~\cite{gsafeguard}                  & SS         & MP             & OA, EN & EC & 02/2025\\
      Wang et al.~\cite{wang2025talk}                  & SS          & MP             & OA, EN & EC & 02/2025\\
      Flow-of-Action~\cite{flow-of-action}                  & SS          & MP, SA            & OA, EN & EC, IC & 02/2025\\
      Yue et al.~\cite{shengbinyue2025multi}                  & SS         & MP             & SE, OA, EN, HU & EC, IC & 02/2025\\
      CODESIM~\cite{codesim}                  & OO         & MP             & SE, OA, EN & EC & 02/2025\\
      LogiAgent~\cite{logiagent}                  & OO          & MP, BB             & SE, OA, EN & EC & 03/2025\\
      GameChat~\cite{gamechat}                  & SS          & SA            & SE, OA & EC & 03/2025\\
      MDTeamGPT~\cite{mdteamgpt}                  & SS          & MP             & SE, OA, EN, HU & EC & 03/2025\\
      AgentFM~\cite{agentfm}                  & OO, ST           & MP, SA             & OA, EN & EC & 04/2025\\
      AutoHMA-LLM~\cite{autohma}                  & OO         & MP            & OA, EN & EC & 04/2025\\
      FactGuard~\cite{factguard}                  & OO, ST           & MP            & OA, EN & EC & 04/2025\\
      Thematic-LM~\cite{thematiclM}                  & OO, ST           & MP             & OA, EN & EC & 04/2025\\
      AutoData~\cite{autodata}                  & OO, ST          & MP, SA             & OA, EN & EC & 05/2025\\
      EcoLang~\cite{ecolang}                  & ST          & MP, BB             & OA, EN & EC, IC & 05/2025\\
      Ronanki et al.~\cite{ronanki2025facilitating}                  & OO, ST          & MP      & OA, HU & EC & 05/2025\\
      Zhang et al.~\cite{zhang2025facilitating}                  & ST          & MP, BB             & OA, EN, HU & EC, IC & 05/2025\\  
      Group Think~\cite{groupthink}                  & OO          & MP             & SE, OA & EC & 05/2025\\     
      Shen et al.~\cite{shen2025optimizing}                  & SS          & MP            & SE, OA, EN, HU & EC, IC & 05/2025\\
      TED~\cite{TED}                  & SS          & MP, BB             & OA, EN & EC, IC & 05/2025\\

      \hline
    \end{tabularx}%

\end{table*}

\subsection{Communication Strategy}\label{subsec:strategie}
\begin{figure}[htbp]
    \centering
    \includegraphics[width=\columnwidth]{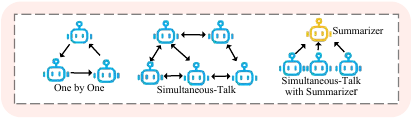} 
    \caption{The Communication Strategies of LLM-MAS} 
    \label{fig:strategie} 
\end{figure}

Communication strategies dictate how and when agents interact within an LLM-based multi-agent system. These strategies govern the timing, sequence, and structure of communication exchanges, significantly affecting system coherence, information accuracy, and the overall effectiveness of agent collaboration. As shown in \cref{fig:strategie}, three widely adopted communication strategies are explored.

\subsubsection{One-by-One}\label{para:stra_onebyone} 
In the One-by-One communication strategy, agents communicate sequentially, each agent responding after fully processing and integrating all previous messages. This structured, turn-based approach maintains a clear context throughout interactions, significantly reducing misunderstandings and ensuring cohesive exchanges. The One-by-One strategy is particularly beneficial for tasks requiring methodical reasoning and detailed integration of inputs from multiple agents.

A typical example is Chain-of-Agents ~\cite{chain_of_agents}, which applies this strategy to manage lengthy text-generation tasks. Each agent summarizes previous contributions before providing its input, efficiently handling contexts beyond typical token limits.

Nevertheless, the strict serial dependency inherent in One-by-One communication imposes two key limitations as the agent population grows. First, latency scales linearly with the number of turns, because each agent must await the full completion of all preceding exchanges before contributing. Second, since inaccuracies introduced by early agents will unfold in each subsequent step, cumulative error propagation becomes more pronounced, posing a risk of semantic drift.

\subsubsection{Simultaneous-Talk}\label{para:stra_aimu} 
The Simultaneous-Talk strategy allows agents to communicate concurrently, without waiting for turns. This strategy fosters a dynamic exchange environment conducive to rapid idea generation and parallel problem-solving. Simultaneous-Talk effectively leverages the diverse perspectives of multiple agents simultaneously, significantly enhancing brainstorming capabilities and uncovering creative solutions.

For instance, Autoagents ~\cite{autoagents} deploys multiple agents to generate ideas or solution paths concurrently. This parallel interaction model quickly reveals diverse possibilities, enhancing creativity and solution robustness. A similar pattern appears in EconAgent~\cite{econagent}, where large-scale agents jointly conduct macroeconomic simulations.

Despite its advantages in agility and diversity, the Simultaneous-Talk approach introduces two key challenges. First, state staleness: without fine-grained synchronisation, agents may base their reasoning on outdated peer contributions, leading to logically inconsistent or redundant outputs. Second, conflict resolution: parallel proposals often overlap or contradict each other, necessitating an additional arbitration layer. Therefore, under this strategy, effective timestamps and lightweight consensus mechanisms are crucial.

\subsubsection{Simultaneous-Talk-with-Summarizer}\label{para:stra_simu_summarizer} 
To address the challenges inherent in pure Simultaneous-Talk strategies, the Simultaneous-Talk with Summarizer strategy integrates a summarizing agent responsible for consolidating concurrent communications into coherent summaries. This summarizer periodically synthesizes inputs from multiple agents into concise, unified messages, ensuring that all agents remain aligned with the overall system context.

The hierarchical and team-based coordination scenarios, exemplified by CausalGPT~\cite{casualgpt_reasoning} and AgentCoord~\cite{agentcoord}, utilize summarizers to aggregate communication outcomes, facilitating clear decision-making and task delegation.

Although introducing a summarizer agent effectively mitigates synchronization issues, it reintroduces some sequential dependencies. Additionally, summarizer agents must accurately interpret and summarize agent communications to avoid misrepresentations or hallucinations, which could compromise the effectiveness of the entire system.

In summary, selecting an appropriate communication strategy depends on the specific context, objectives, and constraints of the multi-agent system. Whether employing structured sequential interactions, parallel exchanges, or hybrid models, each strategy provides distinct advantages and trade-offs, critically influencing the efficiency, responsiveness, and robustness of agent communications.

\subsection{Communication Paradigm}\label{subsec:paradigm}
\begin{figure}[htbp]
    \centering
    \includegraphics[width=\columnwidth]{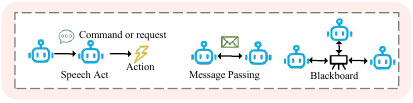} 
    \caption{The Communication Paradigms of LLM-MAS} 
    \label{fig:paradigm} 
\end{figure}

Communication paradigms define how information is represented, transmitted, and interpreted among agents in LLM-based multi-agent systems. By utilizing advanced natural language processing capabilities of LLMs, these paradigms enable agents to conduct richer, more context-sensitive interactions than traditional symbolic communication methods. Drawing on the taxonomy proposed by ~\cite{mas_a_survey}, as shown in \cref{fig:paradigm}, three common communication paradigms are distinguished.

\subsubsection{Message Passing}\label{para:paradigm_messagepassing} 

The Message Passing paradigm refers to direct point-to-point or broadcast communication, in which agents explicitly exchange messages containing information, instructions, or requests. These messages typically consist of natural language content augmented with contextual and reasoning-related information, thereby facilitating precise and efficient information dissemination across agents.

A prime example is the simulation of opinion dynamics using networks of LLM-based agents by Chuang et al.~\cite{simulating_opinion_dynamic}. In this scenario, agents use direct messaging to continuously update each other about evolving opinions, thereby maintaining accurate awareness of the collective state and enabling rapid consensus building. Another notable application is GameChat~\cite{gamechat}, where multiple LLM agents interact via direct message exchanges to achieve safe and socially optimal navigation in constrained environments. 

While effective in ensuring clarity and immediacy of interactions, message passing requires robust message handling systems to manage potential overloads and ensure consistency.

\subsubsection{Speech Act}\label{para:paradigm_speechact} 
The Speech Act paradigm stems from the notion that language serves not only as a medium for information exchange but also as a tool to perform actions. The Speech Act paradigm conceptualizes communication as performative actions where agent utterances are designed to trigger specific actions or state changes within the system. Speech acts typically include instructive, persuasive, or directive components explicitly crafted to influence recipient agent behaviors, dynamically shaping system dynamics based on evolving interaction contexts.

Concretely, speech acts enable agents to steer interactions in contexts that require on-the-fly adaptation, including diplomatic negotiation \cite{polca_mas_for_political,richeliey_diplomacy_society}, collaborative reasoning or debate-driven optimization \cite{debate_improve_llm_1}, and large-scale software co-engineering workflows \cite{chatdev_software_development}. By embedding directive, commissive, or persuasive components into messages, agents can negotiate task allocations, align beliefs, and synchronously coordinate temporal commitments, thereby shaping macro-level system dynamics as the dialogue unfolds.

Yet this expressive power comes at a cost. Ambiguous force-marking, underspecified contextual frames, or divergent belief models can cause illocutionary misfires: the recipient may infer an unintended goal or act on incomplete preconditions, leading to deadlocks or cascading errors. Mitigating such risks calls for standardized performative taxonomies and explicit grounding protocols—possibly augmented with uncertainty-aware LLM decoding—that allow agents to confirm or renegotiate semantic commitments before executing high-impact actions.

\subsubsection{Blackboard}\label{para:paradigm_blackboard} 

The Blackboard paradigm employs a centralized information repository, where agents collaboratively share, retrieve, and coordinate through published messages or updates. This shared medium acts as a communal workspace, significantly enhancing coordination effectiveness and facilitating a unified understanding among agents about current system states, tasks, or decisions. 

This paradigm particularly suits highly coordinated systems like collaborative decision-making or distributed problem-solving, where agents require a shared information pool to align strategies and actions. A typical application example is MetaGPT~\cite{metagpt}, which effectively leverages the blackboard paradigm by implementing a centralized communication repository where software development agents collaboratively post status updates, code snippets, or issue resolutions. This centralization enables efficient information sharing, coordinated task allocation, and rapid consensus formation, streamlining collaborative development efforts. Similarly, MDTeamGPT~\cite{mdteamgpt} demonstrates blackboard usage in medical consultations, where a shared information space ensures medical teams consistently access relevant patient information, diagnostic suggestions, and treatment plans.

While highly beneficial for enhancing information consistency and accessibility, blackboard systems require robust mechanisms to prevent bottlenecks, manage access permissions, and mitigate risks from misinformation or malicious inputs.

In summary, selecting appropriate communication paradigms is crucial for maximizing interaction effectiveness in LLM-based multi-agent systems. Each paradigm offers distinct advantages and considerations, tailored to specific interaction scenarios and system requirements. The choice of paradigm significantly influences overall communication coherence, system responsiveness, and the efficacy of collaborative outcomes.

\subsection{Communication Object}\label{subsec:object}
\begin{figure}[htbp]
    \centering
    \includegraphics[width=\columnwidth]{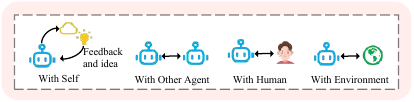} 
    \caption{The Communication Objects of LLM-MAS} 
    \label{fig:object} 
\end{figure}

Communication objects define the entities or targets with which agents interact within an LLM-based multi-agent system. These interactions significantly shape agents' perceptions, decisions, and behaviors. Identifying and clearly defining these communication objects enhances system clarity, enables more targeted interactions, and improves overall coordination and efficiency. As shown in \cref{fig:object}, this section examines four primary communication objects.

\subsubsection{Communication with Self}\label{para:object_self} 
Communication with self refers to internal dialogues or reflective processes within agents, enabling them to deliberate, plan, and refine decisions independently~\cite{react}. This self-interaction is crucial for agents to update their internal state, manage cognitive resources, and facilitate sophisticated reasoning processes that emulate human-like introspection and problem-solving.

For instance, AgentCoord~\cite{agentcoord} uses self-communication to iteratively evaluate and refine coordination strategies among multiple agents. Agents internally reflect on past interactions, hypothesize outcomes of potential actions, and update their strategies accordingly. Similarly, FixAgent~\cite{fixagent_mas_for_debug} utilizes self-reflection during debugging tasks, allowing individual agents to internally assess debugging strategies before implementing solutions, thereby significantly enhancing debugging effectiveness and resource management.

However, internal communication requires robust cognitive frameworks within agents to accurately simulate reflective processes, posing challenges in accurately managing agent cognition.

\subsubsection{Communication with Other Agents}\label{para:object_agent} 

Communication with other agents involves direct interactions among agents, encompassing exchanges of information, coordination of tasks, or negotiation processes, which exist in the majority of LLM-MAS. Effective inter-agent communication is critical for collaborative problem-solving, conflict resolution, and coordinated decision-making.

MAGIS~\cite{magis_mas_for_github} exemplifies robust inter-agent communication in collaborative software development tasks, where developers, testers, and administrators frequently exchange task-specific messages to streamline software issue resolution. Similarly, AgentFM~\cite{agentfm} utilizes structured communication among agents managing database systems, ensuring coordinated responses to failure management and improving system reliability.

Almost all LLM-MAS scenarios involve communication with other agents. Effective inter-agent communication requires sophisticated protocols and mechanisms to ensure synchronization, consistency, and conflict resolution, particularly in large-scale multi-agent systems.

\subsubsection{Communication with Environment}\label{para:object_environment} 
Communication with the environment refers to an agent’s ability to sense external stimuli and adjust its behavior in real time. Unlike communicating with agents or humans, this modality hinges on parsing raw sensor streams including images, audio, proprioception, or simulator state and converting them into actionable knowledge.

EmbodiedGPT~\cite{embodied_agents} demonstrates environmental communication through multimodal perception, where agents process and respond to sensory inputs like visual and audio data, effectively navigating and interacting within dynamic environments. Similarly, ChatSim~\cite{chatsim_mas_make_scene} incorporates environmental feedback into agent interactions during autonomous driving simulations, enabling agents to refine their behaviors based on real-time environmental changes and conditions.

Robust environment communication thus demands accurate, low-latency sensor fusion, continual grounding of textual reasoning in non-linguistic data, and safety guards against perception errors or out-of-distribution inputs. Addressing these challenges is essential for deploying LLM-MAS in dynamic, high-stakes settings such as embodied assistance, industrial automation, or intelligent transportation.

\subsubsection{Communication with Human}\label{para:object_human} 
Communication with humans involves direct interactions between agents and human users or participants. This interaction layer adds complexity to multi-agent systems by requiring agents to understand, interpret, and appropriately respond to human-generated inputs, commands, or feedback, maintaining effective and natural interactions.

PeerGPT~\cite{peergpt} effectively illustrates human-agent communication in collaborative educational scenarios, where agents function as moderators or participants interacting dynamically with children. Agents interpret children's verbal inputs and physical actions, continuously refining their responses to enhance learning outcomes and maintain engagement. Additionally, MedAgents~\cite{medagents} exemplifies human-agent interaction in medical consultations, where agents collaboratively interact with medical professionals to support diagnosis, treatment planning, and patient care coordination. 

These applications highlight three challenges: robust natural-language understanding across diverse user populations, integration of affective and situational signals to maintain rapport, and stringent safety and ethical requirements concerning privacy, bias, and accountability. Addressing these challenges is critical for deploying LLM-MAS that can communicate with humans in a trustworthy and effective manner.

In summary, clearly defining and managing communication objects is essential for optimizing agent interactions in LLM-MAS. Effective management of these interactions significantly influences the system’s adaptability, responsiveness, and overall operational effectiveness.

\subsection{Communication Content}\label{subsec:content}

\begin{figure}[htbp]
    \centering
    \includegraphics[width=\columnwidth]{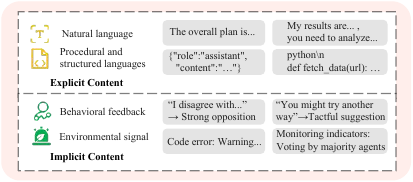} 
    \caption{The Communication Contents of LLM-MAS} 
    \label{fig:content} 
\end{figure}

Communication content specifies the type and nature of information exchanged between agents, influencing their interactions, understanding, and subsequent actions. As shown in \cref{fig:content}, this section categorizes communication content into Explicit and Implicit forms, with further subdivisions to enhance clarity and provide detailed insights into their application contexts.

\subsubsection{Explicit communication}\label{para:content_explicit} 

Explicit communication entails direct information exchange with clearly defined and easily interpretable meaning. Explicit communication can be divided into two forms:

\paragraph{Natural Language.}
Explicit communication in the form of natural language involves direct verbal exchanges articulated in human-readable text or spoken formats. This type of communication leverages the advanced natural language processing capabilities of LLMs, allowing agents to communicate clearly, flexibly, and contextually.

For example, generative agents in sandbox environments~\cite{stanf_villege} frequently utilize natural language communication to negotiate, plan, or resolve conflicts, mirroring complex human interactions. Moreover, agents in platforms like AgentCoord~\cite{agentcoord} employ natural language to articulate strategic decisions clearly, ensuring smooth and effective coordination among agents. The flexibility and expressiveness of natural language enhance the quality of interactions but require sophisticated parsing capabilities to avoid ambiguities or misunderstandings.

\paragraph{Code and Structured Data.}
Communication through code and structured data involves precise, unambiguous exchanges, typically formatted in standardized data structures or procedural languages. This form of explicit communication is essential for tasks requiring high accuracy and clarity, such as detailed instructions, algorithmic descriptions, or structured data exchanges.

For instance, MAGIS~\cite{magis_mas_for_github} utilizes structured code and data exchanges among software development agents to streamline issue tracking and code management processes. Similarly, AutoData~\cite{autodata} applies structured data communication for open web data collection, ensuring efficient and accurate data aggregation and analysis. Structured communication effectively minimizes misinterpretations but demands strict adherence to predefined data standards and formats.

\subsubsection{Implicit communication.}\label{para:content_implicit}

Implicit communication occurs when agents convey information indirectly, through their actions or environmental cues, rather than through explicit statements. Implicit communication relies on agents interpreting contextual cues or feedback. Implicit communication can also be divided into two forms:

\paragraph{Behavioral Feedback.}
Implicit communication through behavioral feedback involves indirect information exchanges inferred from agent actions or interactions. Instead of explicit verbal cues, agents communicate intentions or feedback implicitly through observable behaviors, strategy shifts, or adaptive responses. Such signals can convey intent, commitment, or private information while preserving strategic stealth.

In diplomatic simulations like Richelieu~\cite{richeliey_diplomacy_society}, agent behaviors implicitly communicate negotiation positions and strategies, influencing the negotiation outcomes indirectly. Similarly, Flooding simulations~\cite{community_knowledge_flooding} demonstrate implicit behavioral feedback, where agent strategies dynamically adjust based on the evolving behaviors of peers, effectively communicating strategic adjustments without explicit verbalization. Although powerful, behavioral feedback requires accurate interpretation mechanisms to ensure the correct inference of implicit signals, otherwise it may cause misjudgment or over-interpretation, resulting in impaired system function.

\paragraph{Environmental Signal.}

Environmental signals serve as another implicit communication channel, where changes or conditions in the operational context indirectly influence agent decisions and behaviors. Agents interpret and respond to these signals to dynamically adapt their actions.

For example, ChatSim~\cite{chatsim_mas_make_scene} integrates environmental signals such as real-time traffic or weather conditions in autonomous driving simulations, guiding agent navigation decisions implicitly. Likewise, agents in EcoLANG~\cite{ecolang} interpret economic indicators and environmental signals to adjust consumption and investment behaviors dynamically. Effective utilization of environmental signals depends heavily on the agents' sensitivity and responsiveness to subtle contextual changes.

In summary, clearly distinguishing and effectively managing explicit and implicit communication content is crucial for enhancing interaction accuracy, adaptability, and overall efficacy in LLM-MAS.

\section{Challenges and Opportunities}\label{sec:challenges}
As LLM-MAS continue to garner increasing attention in both research and application domains, their further development faces several significant challenges and emerging opportunities. We analyze several key challenges and research directions based on the article's content.
\subsection{Optimizing the System Design}\label{subsec:opt}
The communication architecture is the foundational component of LLM-MAS. As task complexity increases, traditional communication architectures may no longer suffice. Therefore, the design of hybrid architectures is expected to be a key focus of future research. With more complex structures, the number of agents increases, leading to greater demands on computational resources. Therefore, a major challenge lies in developing communication paradigms that are both efficient and scalable. Meanwhile, how to optimize the allocation of computational resources also needs to be considered. Concurrently, the increasing volume of internal system information poses another challenge. Ensuring that agents correctly interpret and understand this information, while minimizing the risk of hallucinations or misunderstandings, will be a crucial area of investigation.
\subsection{Advancing Research on Agent Competition}\label{subsec:competition}
In a competitive environment, agents can develop more complex strategies, improve decision-making, and promote innovative behaviors by employing techniques such as game theory. However, a key challenge lies in balancing competition and cooperation, as excessive competition may lead to inefficiency or instability. Future research can focus on finding the optimal balance between competition and cooperation, developing scalable competition strategies, and exploring how to safely and effectively integrate competition into real-world applications.

\subsection{Unified Communication Protocol}\label{subsec:unified_protocol}
The rapid emergence and deployment of multiple new communication protocols such as MCP, A2A, ANP, ACP, AITP, and AConP in LLM-MAS underscore the field's dynamism and growth. However, this proliferation also brings critical challenges. One significant issue is functional redundancy, as different protocols often overlap in terms of capabilities such as secure communication, context management, and agent discovery. This redundancy can lead to unnecessary complexity, resource wastage, and increased difficulty in protocol management.

Moreover, the lack of interoperability among existing protocols poses substantial barriers. Different agent groups employing distinct protocols cannot seamlessly communicate or collaborate, significantly hindering the scalability and integration of multi-agent systems. This situation mirrors early-stage internet communication challenges, which were eventually resolved through standardized protocols like HTTPS.

Therefore, the development and adoption of a unified, standardized communication protocol for LLM-MAS is imperative. Such a protocol would provide foundational interoperability, enhance security, simplify integration, and significantly reduce the complexity inherent in managing multiple disparate systems. By achieving a consensus-driven standard akin to HTTPS, LLM-MAS can more effectively harness the collective intelligence and collaborative potential of agents, thereby driving further innovation, reliability, and widespread adoption across diverse application domains.

\subsection{Multimodal Communication}\label{subsec:multimodal}

With the development of large multimodal models, agents in LLM-MAS should not be limited to text-based communication. Communication of multimodal content (text, images, audio, and video) should also be considered.This expansion into multimodal content enables more natural and context-aware interactions, thereby enhancing agents' adaptability and decision-making capabilities. However, there are some challenges to integrating multimodal content. A major issue is how to effectively present and coordinate different modalities in a coherent way that is comprehensible to all agents. In addition, agents not only have to process these different modalities, but also communicate them effectively to one another. Future research should focus on improving the fusion of multimodal data and designing stronger agents in key components for handling multimodal content. 

\subsection{Communication Security}\label{subsec:security}

As LLM-MAS migrate into safety-critical domains, safeguarding the confidentiality, integrity, and authenticity of inter-agent messages is indispensable. Recent studies ~\cite{redteaming,corba,community_knowledge_flooding} demonstrate that adversaries can mount man-in-the-middle, data-tampering, and spoofing attacks, resulting in damage cascading rapidly through the system. Hence, defence techniques originally designed for standalone LLMs ~\cite{defense1,defense2} must be embedded within the communication layer itself. Preventing eavesdropping and forgery in decentralized, dynamic topologies where agents may join or leave at will requires cryptographically grounded protocols that combine end-to-end encryption, fine-grained authentication, adaptive key management, and provenance tracking. Developing such adaptive secure-communication frameworks will be pivotal for deploying LLM-MAS in high-stakes settings such as cooperative autonomous driving and clinical decision support, where even transient message corruption can have cascading real-world consequences.

\subsection{Benchmarks and Evaluation}\label{subsec:benchmark}
The rapid diffusion of LLM-MAS across software engineering, game-playing, and social-simulation tasks has outpaced the development of rigorous evaluation suites. Existing datasets such as MultiAgentBench and RealWorldBench~\cite{multiagentbench,realworldbench} cover a handful of cooperative scenarios, but fall short of spanning heterogeneous domains, interaction paradigms, or agent-population scales.

Current leaderboards also remain agent-centric: they report per-agent task accuracy or reward but rarely capture system-level properties, including coordination efficiency, communication bandwidth and latency, robustness to stale or deceptive messages, and group-level fairness and safety. Without these dimensions, evaluations cannot distinguish architectures that merely aggregate strong single agents from those that yield genuinely emergent collaboration.

We therefore advocate a next-generation benchmark suite that spans cooperation, competition, and mixed-motive settings and reports a multi-granular metric set from individual competence to collective utility to provide a holistic portrait of LLM-MAS performance.
\section{Conclusion}
In this survey, we have presented a communication-centric framework for understanding LLM-MAS, emphasizing how effective communication shapes system performance. By distinguishing between system-level communication and system-internal communication, we systematically explored critical components, including architectures, communication goals, protocols, strategies, paradigms, objects, and content. The significant impacts of these components on the efficiency, scalability, and adaptability of LLM-MAS are discussed in detail. We also discuss challenges and future opportunities. Key challenges identified include interoperability, multimodal integration, and security. Future research directions should focus on addressing these issues, optimizing system designs, developing unified protocols, and establishing comprehensive benchmarks. We expect this communication-centric perspective to inform and inspire ongoing research, driving further development of LLM-MAS.

\section*{Acknowledgement}
This research is supported by the National Natural Science Foundation of China (No.62272025, No.U22B2021 and
No.62362048)

\section*{Competing Interest}
The authors declare that they have no competing interests or financial conflicts to disclose.

\bibliographystyle{fcs}
\bibliography{ref}

@article{2,
  title   = {Wolfhard {H G}. Flames. 2nd ed},
  author  = {Gaydon, A G},
  journal = {London: Chapman and Hall Ltd},
  year    = {1960}
}

@inproceedings{3,
  author    = {Karger, David R. and Ruhl, Matthias},
  title     = {Simple efficient load balancing algorithms for peer-to-peer systems},
  booktitle = {Proceedings of the sixteenth annual ACM symposium on parallelism in algorithms and architectures},
  series    = {SPAA '04},
  year      = {2004},
  pages     = {36--43}
}

@inbook{4,
  chapter   = {Computational complexity},
  pages     = {260-265},
  title     = {Encyclopedia of Computer Science},
  publisher = {Wiley},
  year      = {1994},
  editor    = {Anthony Ralston and Edwin D. Reilly and David Hemmendinger},
  author    = {Christos H. Papadimitriou},
  address   = {Chichester}
}

@article{llm_agent_define,
  title={A survey on large language model based autonomous agents},
  author={Wang, Lei and Ma, Chen and Feng, Xueyang and Zhang, Zeyu and Yang, Hao and Zhang, Jingsen and Chen, Zhiyuan and Tang, Jiakai and Chen, Xu and Lin, Yankai and others},
  journal={Frontiers of Computer Science},
  volume={18},
  number={6},
  pages={186345},
  year={2024},
  publisher={Springer}
}

@inproceedings{react,
  title={React: Synergizing reasoning and acting in language models},
  author={Yao, Shunyu and Zhao, Jeffrey and Yu, Dian and Du, Nan and Shafran, Izhak and Narasimhan, Karthik and Cao, Yuan},
  booktitle={International Conference on Learning Representations (ICLR)},
  year={2023}
}

@article{rag_1,
  title={A Survey of Graph Retrieval-Augmented Generation for Customized Large Language Models},
  author={Zhang, Qinggang and Chen, Shengyuan and Bei, Yuanchen and Yuan, Zheng and Zhou, Huachi and Hong, Zijin and Dong, Junnan and Chen, Hao and Chang, Yi and Huang, Xiao},
  journal={arXiv preprint arXiv:2501.13958},
  year={2025}
}

@article{rag_2,
  title={Entity alignment with noisy annotations from large language models},
  author={Chen, Shengyuan and Zhang, Qinggang and Dong, Junnan and Hua, Wen and Li, Qing and Huang, Xiao},
  journal={arXiv preprint arXiv:2405.16806},
  year={2024}
}

@article{single_limit,
  title={Multi-agent collaboration: Harnessing the power of intelligent llm agents},
  author={Talebirad, Yashar and Nadiri, Amirhossein},
  journal={arXiv preprint arXiv:2306.03314},
  year={2023}
}

@article{agent_survey_1,
  title={Exploring large language model based intelligent agents: Definitions, methods, and prospects},
  author={Cheng, Yuheng and Zhang, Ceyao and Zhang, Zhengwen and Meng, Xiangrui and Hong, Sirui and Li, Wenhao and Wang, Zihao and Wang, Zekai and Yin, Feng and Zhao, Junhua and others},
  journal={arXiv preprint arXiv:2401.03428},
  year={2024}
}

@article{mas_survey_1,
  title={A survey on LLM-based multi-agent systems: workflow, infrastructure, and challenges},
  author={Li, Xinyi and Wang, Sai and Zeng, Siqi and Wu, Yu and Yang, Yi},
  journal={Vicinagearth},
  volume={1},
  number={1},
  pages={9},
  year={2024},
  publisher={Springer}
}

@article{mas_survey_2,
  title={Large language model based multi-agents: A survey of progress and challenges},
  author={Guo, Taicheng and Chen, Xiuying and Wang, Yaqi and Chang, Ruidi and Pei, Shichao and Chawla, Nitesh V and Wiest, Olaf and Zhang, Xiangliang},
  journal={arXiv preprint arXiv:2402.01680},
  year={2024}
}

@article{mas_survey_3,
  title={LLM multi-agent systems: Challenges and open problems},
  author={Han, Shanshan and Zhang, Qifan and Yao, Yuhang and Jin, Weizhao and Xu, Zhaozhuo and He, Chaoyang},
  journal={arXiv preprint arXiv:2402.03578},
  year={2024}
}

@article{mas_application_survey_simulation,
  title={Large language models empowered agent-based modeling and simulation: A survey and perspectives},
  author={Gao, Chen and Lan, Xiaochong and Li, Nian and Yuan, Yuan and Ding, Jingtao and Zhou, Zhilun and Xu, Fengli and Li, Yong},
  journal={Humanities and Social Sciences Communications},
  volume={11},
  number={1},
  pages={1--24},
  year={2024},
  publisher={Palgrave}
}

@article{mas_application_survey_1,
  title={From llms to llm-based agents for software engineering: A survey of current, challenges and future},
  author={Jin, Haolin and Huang, Linghan and Cai, Haipeng and Yan, Jun and Li, Bo and Chen, Huaming},
  journal={arXiv preprint arXiv:2408.02479},
  year={2024}
}

@article{mas_application_survey_2,
  title={Large language model-based agents for software engineering: A survey},
  author={Liu, Junwei and Wang, Kaixin and Chen, Yixuan and Peng, Xin and Chen, Zhenpeng and Zhang, Lingming and Lou, Yiling},
  journal={arXiv preprint arXiv:2409.02977},
  year={2024}
}

@article{communication_1,
  title={A mathematical theory of communication},
  author={Shannon, Claude E},
  journal={The Bell system technical journal},
  volume={27},
  number={3},
  pages={379--423},
  year={1948},
  publisher={Nokia Bell Labs}
}

@article{communication_2,
  title={The contract net protocol: High-level communication and control in a distributed problem solver},
  author={Smith, Reid G},
  journal={IEEE Transactions on computers},
  volume={29},
  number={12},
  pages={1104--1113},
  year={1980},
  publisher={IEEE Computer Society}
}

@article{embodied_agents,
  title={Embodiedgpt: Vision-language pre-training via embodied chain of thought},
  author={Mu, Yao and Zhang, Qinglong and Hu, Mengkang and Wang, Wenhai and Ding, Mingyu and Jin, Jun and Wang, Bin and Dai, Jifeng and Qiao, Yu and Luo, Ping},
  journal={Advances in Neural Information Processing Systems},
  volume={36},
  pages={25081--25094},
  year={2023}
}

@article{mas_define,
  title={A decentralized cluster formation containment framework for multirobot systems},
  author={Hu, Junyan and Bhowmick, Parijat and Jang, Inmo and Arvin, Farshad and Lanzon, Alexander},
  journal={IEEE Transactions on Robotics},
  volume={37},
  number={6},
  pages={1936--1955},
  year={2021},
  publisher={IEEE}
}

@article{mas_a_survey,
  title={Multi-agent systems: A survey},
  author={Dorri, Ali and Kanhere, Salil S and Jurdak, Raja},
  journal={Ieee Access},
  volume={6},
  pages={28573--28593},
  year={2018},
  publisher={IEEE}
}

@article{boostrapping,
  title={Bootstrapping llm-based task-oriented dialogue agents via self-talk},
  author={Ulmer, Dennis and Mansimov, Elman and Lin, Kaixiang and Sun, Justin and Gao, Xibin and Zhang, Yi},
  journal={arXiv preprint arXiv:2401.05033},
  year={2024}
}

@article{dataset_generation,
  title={Synthetic dialogue dataset generation using llm agents},
  author={Abdullin, Yelaman and Molla-Aliod, Diego and Ofoghi, Bahadorreza and Yearwood, John and Li, Qingyang},
  journal={arXiv preprint arXiv:2401.17461},
  year={2024}
}

@inproceedings{debate_improve_llm_1,
  title={Improving factuality and reasoning in language models through multiagent debate},
  author={Du, Yilun and Li, Shuang and Torralba, Antonio and Tenenbaum, Joshua B and Mordatch, Igor},
  booktitle={Forty-first International Conference on Machine Learning},
  year={2023}
}

@article{evaluating_llm_game_ability,
  title={How Far Are We on the Decision-Making of LLMs? Evaluating LLMs' Gaming Ability in Multi-Agent Environments},
  author={Huang, Jen-tse and Li, Eric John and Lam, Man Ho and Liang, Tian and Wang, Wenxuan and Yuan, Youliang and Jiao, Wenxiang and Wang, Xing and Tu, Zhaopeng and Lyu, Michael R},
  journal={arXiv preprint arXiv:2403.11807},
  year={2024}
}

@article{simulating_opinion_dynamic,
  title={Simulating opinion dynamics with networks of llm-based agents},
  author={Chuang, Yun-Shiuan and Goyal, Agam and Harlalka, Nikunj and Suresh, Siddharth and Hawkins, Robert and Yang, Sijia and Shah, Dhavan and Hu, Junjie and Rogers, Timothy T},
  journal={arXiv preprint arXiv:2311.09618},
  year={2023}
}

@article{chatdev_software_development,
  title={Chatdev: Communicative agents for software development},
  author={Qian, Chen and Liu, Wei and Liu, Hongzhang and Chen, Nuo and Dang, Yufan and Li, Jiahao and Yang, Cheng and Chen, Weize and Su, Yusheng and Cong, Xin and others},
  journal={arXiv preprint arXiv:2307.07924},
  year={2023}
}

@article{autodefense_against_jailbreak,
  title={Autodefense: Multi-agent llm defense against jailbreak attacks},
  author={Zeng, Yifan and Wu, Yiran and Zhang, Xiao and Wang, Huazheng and Wu, Qingyun},
  journal={arXiv preprint arXiv:2403.04783},
  year={2024}
}

@article{casualgpt_reasoning,
  title={Towards causalgpt: A multi-agent approach for faithful knowledge reasoning via promoting causal consistency in llms},
  author={Tang, Ziyi and Wang, Ruilin and Chen, Weixing and Wang, Keze and Liu, Yang and Chen, Tianshui and Lin, Liang},
  journal={arXiv preprint arXiv:2308.11914},
  year={2023}
}

@article{soa_code_generation,
  title={Self-organized agents: A llm multi-agent framework toward ultra large-scale code generation and optimization},
  author={Ishibashi, Yoichi and Nishimura, Yoshimasa},
  journal={arXiv preprint arXiv:2404.02183},
  year={2024}
}

@article{fincon_decision_making,
  title={Fincon: A synthesized llm multi-agent system with conceptual verbal reinforcement for enhanced financial decision making},
  author={Yu, Yangyang and Yao, Zhiyuan and Li, Haohang and Deng, Zhiyang and Jiang, Yuechen and Cao, Yupeng and Chen, Zhi and Suchow, Jordan and Cui, Zhenyu and Liu, Rong and others},
  journal={Advances in Neural Information Processing Systems},
  volume={37},
  pages={137010--137045},
  year={2024}
}

@article{mas_benchmark,
  title={Benchmark self-evolving: A multi-agent framework for dynamic llm evaluation},
  author={Wang, Siyuan and Long, Zhuohan and Fan, Zhihao and Wei, Zhongyu and Huang, Xuanjing},
  journal={arXiv preprint arXiv:2402.11443},
  year={2024}
}

@INPROCEEDINGS{social_media_regulation,
  author={Cai, Jinyu and Li, Jialong and Zhang, Mingyue and Li, Munan and Wang, Chen-Shu and Tei, Kenji},
  booktitle={2024 IEEE Congress on Evolutionary Computation (CEC)}, 
  title={Language Evolution for Evading Social Media Regulation via LLM-Based Multi-Agent Simulation}, 
  year={2024},
  volume={},
  number={},
  pages={1-10},
  doi={10.1109/CEC60901.2024.10612015}}

@article{community_knowledge_flooding,
  title={Flooding spread of manipulated knowledge in llm-based multi-agent communities},
  author={Ju, Tianjie and Wang, Yiting and Ma, Xinbei and Cheng, Pengzhou and Zhao, Haodong and Wang, Yulong and Liu, Lifeng and Xie, Jian and Zhang, Zhuosheng and Liu, Gongshen},
  journal={arXiv preprint arXiv:2407.07791},
  year={2024}
}

@article{mas_for_jvbensha,
  title={Deciphering digital detectives: Understanding llm behaviors and capabilities in multi-agent mystery games},
  author={Wu, Dekun and Shi, Haochen and Sun, Zhiyuan and Liu, Bang},
  journal={arXiv preprint arXiv:2312.00746},
  year={2023}
}

@article{metagpt,
  title={Metagpt: Meta programming for multi-agent collaborative framework},
  author={Hong, Sirui and Zheng, Xiawu and Chen, Jonathan and Cheng, Yuheng and Wang, Jinlin and Zhang, Ceyao and Wang, Zili and Yau, Steven Ka Shing and Lin, Zijuan and Zhou, Liyang and others},
  journal={arXiv preprint arXiv:2308.00352},
  volume={3},
  number={4},
  pages={6},
  year={2023}
}

@article{magis_mas_for_github,
  title={Magis: Llm-based multi-agent framework for github issue resolution},
  author={Tao, Wei and Zhou, Yucheng and Wang, Yanlin and Zhang, Wenqiang and Zhang, Hongyu and Cheng, Yu},
  journal={Advances in Neural Information Processing Systems},
  volume={37},
  pages={51963--51993},
  year={2024}
}

@article{agentcoord,
  title={AgentCoord: Visually exploring coordination strategy for llm-based multi-agent collaboration},
  author={Pan, Bo and Lu, Jiaying and Wang, Ke and Zheng, Li and Wen, Zhen and Feng, Yingchaojie and Zhu, Minfeng and Chen, Wei},
  journal={arXiv preprint arXiv:2404.11943},
  year={2024}
}

@article{autoagents,
  title={Autoagents: A framework for automatic agent generation},
  author={Chen, Guangyao and Dong, Siwei and Shu, Yu and Zhang, Ge and Sesay, Jaward and Karlsson, B{\"o}rje F and Fu, Jie and Shi, Yemin},
  journal={arXiv preprint arXiv:2309.17288},
  year={2023}
}

@article{mas_for_software_2,
  title={Experiential co-learning of software-developing agents},
  author={Qian, Chen and Dang, Yufan and Li, Jiahao and Liu, Wei and Xie, Zihao and Wang, Yifei and Chen, Weize and Yang, Cheng and Cong, Xin and Che, Xiaoyin and others},
  journal={arXiv preprint arXiv:2312.17025},
  year={2023}
}

@article{classroom_simulation,
  title={Simulating classroom education with llm-empowered agents},
  author={Zhang, Zheyuan and Zhang-Li, Daniel and Yu, Jifan and Gong, Linlu and Zhou, Jinchang and Hao, Zhanxin and Jiang, Jianxiao and Cao, Jie and Liu, Huiqin and Liu, Zhiyuan and others},
  journal={arXiv preprint arXiv:2406.19226},
  year={2024}
}

@article{polca_mas_for_political,
  title={Modelling political coalition negotiations using llm-based agents},
  author={Moghimifar, Farhad and Li, Yuan-Fang and Thomson, Robert and Haffari, Gholamreza},
  journal={arXiv preprint arXiv:2402.11712},
  year={2024}
}

@inproceedings{recon_thinking,
  title={Boosting llm agents with recursive contemplation for effective deception handling},
  author={Wang, Shenzhi and Liu, Chang and Zheng, Zilong and Qi, Siyuan and Chen, Shuo and Yang, Qisen and Zhao, Andrew and Wang, Chaofei and Song, Shiji and Huang, Gao},
  booktitle={Findings of the Association for Computational Linguistics ACL 2024},
  pages={9909--9953},
  year={2024}
}

@inproceedings{stanf_villege,
  title={Generative agents: Interactive simulacra of human behavior},
  author={Park, Joon Sung and O'Brien, Joseph and Cai, Carrie Jun and Morris, Meredith Ringel and Liang, Percy and Bernstein, Michael S},
  booktitle={Proceedings of the 36th annual acm symposium on user interface software and technology},
  pages={1--22},
  year={2023}
}

@article{govsim,
  title={Cooperate or collapse: Emergence of sustainable cooperation in a society of llm agents},
  author={Piatti, Giorgio and Jin, Zhijing and Kleiman-Weiner, Max and Sch{\"o}lkopf, Bernhard and Sachan, Mrinmaya and Mihalcea, Rada},
  journal={Advances in Neural Information Processing Systems},
  volume={37},
  pages={111715--111759},
  year={2024}
}

@article{sct_society,
  title={Artificial leviathan: Exploring social evolution of llm agents through the lens of hobbesian social contract theory},
  author={Dai, Gordon and Zhang, Weijia and Li, Jinhan and Yang, Siqi and Rao, Srihas and Caetano, Arthur and Sra, Misha and others},
  journal={arXiv preprint arXiv:2406.14373},
  year={2024}
}

@article{econagent,
  title={Econagent: large language model-empowered agents for simulating macroeconomic activities},
  author={Li, Nian and Gao, Chen and Li, Mingyu and Li, Yong and Liao, Qingmin},
  journal={arXiv preprint arXiv:2310.10436},
  year={2023}
}

@article{richeliey_diplomacy_society,
  title={Richelieu: Self-evolving llm-based agents for ai diplomacy},
  author={Guan, Zhenyu and Kong, Xiangyu and Zhong, Fangwei and Wang, Yizhou},
  journal={Advances in Neural Information Processing Systems},
  volume={37},
  pages={123471--123497},
  year={2024}
}

@article{fixagent_mas_for_debug,
  title={A unified debugging approach via llm-based multi-agent synergy},
  author={Lee, Cheryl and Xia, Chunqiu Steven and Yang, Longji and Huang, Jen-tse and Zhu, Zhouruixin and Zhang, Lingming and Lyu, Michael R},
  journal={arXiv preprint arXiv:2404.17153},
  year={2024}
}

@inproceedings{peergpt,
  title={PeerGPT: Probing the Roles of LLM-based Peer Agents as Team Moderators and Participants in Children's Collaborative Learning},
  author={Liu, Jiawen and Yao, Yuanyuan and An, Pengcheng and Wang, Qi},
  booktitle={Extended Abstracts of the CHI Conference on Human Factors in Computing Systems},
  pages={1--6},
  year={2024}
}

@inproceedings{chatsim_mas_make_scene,
  title={Editable scene simulation for autonomous driving via collaborative llm-agents},
  author={Wei, Yuxi and Wang, Zi and Lu, Yifan and Xu, Chenxin and Liu, Changxing and Zhao, Hao and Chen, Siheng and Wang, Yanfeng},
  booktitle={Proceedings of the IEEE/CVF Conference on Computer Vision and Pattern Recognition},
  pages={15077--15087},
  year={2024}
}

@article{mas_use_tool,
  title={Tulip Agent--Enabling LLM-Based Agents to Solve Tasks Using Large Tool Libraries},
  author={Ocker, Felix and Tanneberg, Daniel and Eggert, Julian and Gienger, Michael},
  journal={arXiv preprint arXiv:2407.21778},
  year={2024}
}

@article{agent4debate,
  title={Can LLMs Beat Humans in Debating? A Dynamic Multi-agent Framework for Competitive Debate},
  author={Zhang, Yiqun and Yang, Xiaocui and Feng, Shi and Wang, Daling and Zhang, Yifei and Song, Kaisong},
  journal={arXiv preprint arXiv:2408.04472},
  year={2024}
}

@article{mas_for_poetry_generation,
  title={LLM-based multi-agent poetry generation in non-cooperative environments},
  author={Zhang, Ran and Eger, Steffen},
  journal={arXiv preprint arXiv:2409.03659},
  year={2024}
}

@TechReport{shallwetalk,
  author={Zengqing Wu and Run Peng and Shuyuan Zheng and Qianying Liu and Xu Han and Brian Inhyuk Kwon and Makoto Onizuka and Shaojie Tang and Chuan Xiao},
  title={{Shall We Team Up: Exploring Spontaneous Cooperation of Competing LLM Agents}},
  year=2024,
  month=Feb,
  institution={arXiv.org},
  type={Papers},
  url={https://ideas.repec.org/p/arx/papers/2402.12327.html},
  number={2402.12327},
  doi={},
}

@article{debate_2,
  title={Can llms produce faithful explanations for fact-checking? towards faithful explainable fact-checking via multi-agent debate},
  author={Kim, Kyungha and Lee, Sangyun and Huang, Kung-Hsiang and Chan, Hou Pong and Li, Manling and Ji, Heng},
  journal={arXiv preprint arXiv:2402.07401},
  year={2024}
}

@inproceedings{blockagents,
  title={BlockAgents: Towards Byzantine-Robust LLM-Based Multi-Agent Coordination via Blockchain},
  author={Chen, Bei and Li, Gaolei and Lin, Xi and Wang, Zheng and Li, Jianhua},
  booktitle={Proceedings of the ACM Turing Award Celebration Conference-China 2024},
  pages={187--192},
  year={2024}
}

@article{mas_for_defence_attack,
  title={Learn to Disguise: Avoid Refusal Responses in LLM's Defense via a Multi-agent Attacker-Disguiser Game},
  author={Xu, Qianqiao and Tian, Zhiliang and Wu, Hongyan and Huang, Zhen and Song, Yiping and Liu, Feng and Li, Dongsheng},
  journal={arXiv preprint arXiv:2404.02532},
  year={2024}
}

@article{mas_for_guandan,
  title={Evaluating and enhancing llms agent based on theory of mind in guandan: A multi-player cooperative game under imperfect information},
  author={Yim, Yauwai and Chan, Chunkit and Shi, Tianyu and Deng, Zheye and Fan, Wei and Zheng, Tianshi and Song, Yangqiu},
  journal={arXiv preprint arXiv:2408.02559},
  year={2024}
}

@article{chain_of_agents,
  title={Chain of agents: Large language models collaborating on long-context tasks},
  author={Zhang, Yusen and Sun, Ruoxi and Chen, Yanfei and Pfister, Tomas and Zhang, Rui and Arik, Sercan},
  journal={Advances in Neural Information Processing Systems},
  volume={37},
  pages={132208--132237},
  year={2024}
}

@inproceedings{jd_recommendation_system,
  title={A hybrid multi-agent conversational recommender system with llm and search engine in e-commerce},
  author={Nie, Guangtao and Zhi, Rong and Yan, Xiaofan and Du, Yufan and Zhang, Xiangyang and Chen, Jianwei and Zhou, Mi and Chen, Hongshen and Li, Tianhao and Cheng, Ziguang and others},
  booktitle={Proceedings of the 18th ACM Conference on Recommender Systems},
  pages={745--747},
  year={2024}
}

@misc{mcp,
  author       = {{Model Context Protocol}},
  title        = {Introduction to model context protocol (MCP)},
  howpublished = {\url{https://modelcontextprotocol.io/introduction}},
  year         = {2024},
  note         = {Accessed: May 2025}
}

@misc{a2a,
  author       = {{Google}},
  title        = {Agent2Agent (A2A) Protocol},
  howpublished = {\url{https://google.github.io/A2A/}},
  year         = {2024},
  note         = {Accessed: May 2025}
}

@misc{anp,
  author       = {{Agent Network Protocol Contributors}},
  title        = {Agent Network Protocol Official Website},
  howpublished = {\url{https://agent-network-protocol.com/}},
  year         = {2024},
  note         = {Accessed: May 2025}
}

@misc{acp,
  author       = {{Linux Foundation AI and LBM Data}},
  title        = {ACP: Agent Communication Protocol},
  howpublished = {\url{https://github.com/orgs/i-am-bee/discussions/284}},
  year         = {2025},
  note         = {Accessed: May 2025}
}

@misc{aitp,
  author       = {{NEAR}},
  title        = {AITP: Agent Interaction \& Transaction Protocol},
  howpublished = {\url{https://aitp.dev/}},
  year         = {2025},
  note         = {Accessed: May 2025}
}

@misc{aconp,
  author       = {{Galileo Cisco and Langchain}},
  title        = {Agent Connect Protocol},
  howpublished = {\url{https://spec.acp.agntcy.org/}},
  year         = {2025},
  note         = {Accessed: May 2025}
}

@book{traditionalmas,
  title={An introduction to multiagent systems},
  author={Wooldridge, Michael},
  year={2009},
  publisher={John wiley \& sons}
}

@book{mas_ctde,
  title={A concise introduction to decentralized POMDPs},
  author={Oliehoek, Frans A and Amato, Christopher and others},
  volume={1},
  year={2016},
  publisher={Springer}
}

@article{toolllm,
  title={Toolllm: Facilitating large language models to master 16000+ real-world apis},
  author={Qin, Yujia and Liang, Shihao and Ye, Yining and Zhu, Kunlun and Yan, Lan and Lu, Yaxi and Lin, Yankai and Cong, Xin and Tang, Xiangru and Qian, Bill and others},
  journal={arXiv preprint arXiv:2307.16789},
  year={2023}
}

@article{agentfm,
  title={AgentFM: Role-Aware Failure Management for Distributed Databases with LLM-Driven Multi-Agents},
  author={Zhang, Lingzhe and Zhai, Yunpeng and Jia, Tong and Huang, Xiaosong and Duan, Chiming and Li, Ying},
  journal={arXiv preprint arXiv:2504.06614},
  year={2025}
}

@inproceedings{fan2025llm,
  title={An LLM-based Framework for Biomedical Terminology Normalization in Social Media via Multi-Agent Collaboration},
  author={Fan, Yongqi and Xue, Kui and Li, Zelin and Zhang, Xiaofan and Ruan, Tong},
  booktitle={Proceedings of the 31st International Conference on Computational Linguistics},
  pages={10712--10726},
  year={2025}
}

@article{autodata,
  title={AutoData: A Multi-Agent System for Open Web Data Collection},
  author={Ma, Tianyi and Qian, Yiyue and Zhang, Zheyuan and Wang, Zehong and Qian, Xiaoye and Bai, Feifan and Ding, Yifan and Luo, Xuwei and Zhang, Shinan and Murugesan, Keerthiram and others},
  journal={arXiv preprint arXiv:2505.15859},
  year={2025}
}

@article{autohma,
  title={AutoHMA-LLM: Efficient Task Coordination and Execution in Heterogeneous Multi-Agent Systems Using Hybrid Large Language Models},
  author={Yang, Tingting and Feng, Ping and Guo, Qixin and Zhang, Jindi and Ning, Jiahong and Wang, Xinghan and Mao, Zhongyang},
  journal={IEEE Transactions on Cognitive Communications and Networking},
  year={2025},
  publisher={IEEE}
}

@inproceedings{ma2025communication,
  title={Communication Makes Perfect: Persuasion Dataset Construction via Multi-LLM Communication},
  author={Ma, Weicheng and Zhang, Hefan and Yang, Ivory and Ji, Shiyu and Chen, Joice and Hashemi, Farnoosh and Mohole, Shubham and Gearey, Ethan and Macy, Michael and Hassanpour, Saeed and others},
  booktitle={Proceedings of the 2025 Conference of the Nations of the Americas Chapter of the Association for Computational Linguistics: Human Language Technologies (Volume 1: Long Papers)},
  pages={4017--4045},
  year={2025}
}

@article{ecolang,
  title={EcoLANG: Efficient and Effective Agent Communication Language Induction for Social Simulation},
  author={Mou, Xinyi and Qian, Chen and Liu, Wei and Huang, Xuanjing and Wei, Zhongyu},
  journal={arXiv preprint arXiv:2505.06904},
  year={2025}
}

@article{ronanki2025facilitating,
  title={Facilitating Trustworthy Human-Agent Collaboration in LLM-based Multi-Agent System oriented Software Engineering},
  author={Ronanki, Krishna},
  journal={arXiv preprint arXiv:2505.04251},
  year={2025}
}

@article{zhang2025facilitating,
  title={Facilitating Video Story Interaction with Multi-Agent Collaborative System},
  author={Zhang, Yiwen and Hao, Jianing and Wang, Zhan and Sheng, Hongling and Zeng, Wei},
  journal={arXiv preprint arXiv:2505.03807},
  year={2025}
}

@article{factguard,
  title={FactGuard: Leveraging Multi-Agent Systems to Generate Answerable and Unanswerable Questions for Enhanced Long-Context LLM Extraction},
  author={Zhang, Qian-Wen and Li, Fang and Wang, Jie and Qiao, Lingfeng and Yu, Yifei and Yin, Di and Sun, Xing},
  journal={arXiv preprint arXiv:2504.05607},
  year={2025}
}

@article{flow-of-action,
  title={Flow-of-Action: SOP Enhanced LLM-Based Multi-Agent System for Root Cause Analysis},
  author={Pei, Changhua and Wang, Zexin and Liu, Fengrui and Li, Zeyan and Liu, Yang and He, Xiao and Kang, Rong and Zhang, Tieying and Chen, Jianjun and Li, Jianhui and others},
  journal={arXiv preprint arXiv:2502.08224},
  year={2025}
}

@article{gsafeguard,
  title={G-safeguard: A topology-guided security lens and treatment on llm-based multi-agent systems},
  author={Wang, Shilong and Zhang, Guibin and Yu, Miao and Wan, Guancheng and Meng, Fanci and Guo, Chongye and Wang, Kun and Wang, Yang},
  journal={arXiv preprint arXiv:2502.11127},
  year={2025}
}

@article{gamechat,
  title={GameChat: Multi-LLM Dialogue for Safe, Agile, and Socially Optimal Multi-Agent Navigation in Constrained Environments},
  author={Mahadevan, Vagul and Zhang, Shangtong and Chandra, Rohan},
  journal={arXiv preprint arXiv:2503.12333},
  year={2025}
}

@article{groupthink,
  title={Group Think: Multiple Concurrent Reasoning Agents Collaborating at Token Level Granularity},
  author={Hsu, Chan-Jan and Buffelli, Davide and McGowan, Jamie and Liao, Feng-Ting and Chen, Yi-Chang and Vakili, Sattar and Shiu, Da-shan},
  journal={arXiv preprint arXiv:2505.11107},
  year={2025}
}

@inproceedings{li2025knowledge,
  title={Knowledge tagging with large language model based multi-agent system},
  author={Li, Hang and Xu, Tianlong and Chang, Ethan and Wen, Qingsong},
  booktitle={Proceedings of the AAAI Conference on Artificial Intelligence},
  volume={39},
  number={28},
  pages={28775--28782},
  year={2025}
}

@article{kalyuzhnaya2025llm,
  title={LLM Agents for Smart City Management: Enhancing Decision Support Through Multi-Agent AI Systems.},
  author={Kalyuzhnaya, Anna and Mityagin, Sergey and Lutsenko, Elizaveta and Getmanov, Andrey and Aksenkin, Yaroslav and Fatkhiev, Kamil and Fedorin, Kirill and Nikitin, Nikolay O and Chichkova, Natalia and Vorona, Vladimir and others},
  journal={Smart Cities (2624-6511)},
  volume={8},
  number={1},
  year={2025}
}

@article{wang2025llm,
  title={LLM-powered Multi-agent Framework for Goal-oriented Learning in Intelligent Tutoring System},
  author={Wang, Tianfu and Zhan, Yi and Lian, Jianxun and Hu, Zhengyu and Yuan, Nicholas Jing and Zhang, Qi and Xie, Xing and Xiong, Hui},
  journal={arXiv preprint arXiv:2501.15749},
  year={2025}
}

@article{luo2025llm,
  title={LLM-Powered Multi-Agent System for Automated Crypto Portfolio Management},
  author={Luo, Yichen and Feng, Yebo and Xu, Jiahua and Tasca, Paolo and Liu, Yang},
  journal={arXiv preprint arXiv:2501.00826},
  year={2025}
}

@article{logiagent,
  title={LogiAgent: Automated Logical Testing for REST Systems with LLM-Based Multi-Agents},
  author={Zhang, Ke and Zhang, Chenxi and Wang, Chong and Zhang, Chi and Wu, YaChen and Xing, Zhenchang and Liu, Yang and Li, Qingshan and Peng, Xin},
  journal={arXiv preprint arXiv:2503.15079},
  year={2025}
}

@article{macm,
  title={Macm: Utilizing a multi-agent system for condition mining in solving complex mathematical problems},
  author={Lei, Bin and Zhang, Yi and Zuo, Shan and Payani, Ali and Ding, Caiwen},
  journal={arXiv preprint arXiv:2404.04735},
  year={2024}
}

@article{mdteamgpt,
  title={MDTeamGPT: A Self-Evolving LLM-based Multi-Agent Framework for Multi-Disciplinary Team Medical Consultation},
  author={Chen, Kai and Li, Xinfeng and Yang, Tianpei and Wang, Hewei and Dong, Wei and Gao, Yang},
  journal={arXiv preprint arXiv:2503.13856},
  year={2025}
}

@article{medagents,
  title={Medagents: Large language models as collaborators for zero-shot medical reasoning},
  author={Tang, Xiangru and Zou, Anni and Zhang, Zhuosheng and Li, Ziming and Zhao, Yilun and Zhang, Xingyao and Cohan, Arman and Gerstein, Mark},
  journal={arXiv preprint arXiv:2311.10537},
  year={2023}
}

@inproceedings{shengbinyue2025multi,
  title={Multi-Agent Simulator Drives Language Models for Legal Intensive Interaction},
  author={ShengbinYue, ShengbinYue and Huang, Ting and Jia, Zheng and Wang, Siyuan and Liu, Shujun and Song, Yun and Huang, Xuan-Jing and Wei, Zhongyu},
  booktitle={Findings of the Association for Computational Linguistics: NAACL 2025},
  pages={6537--6570},
  year={2025}
}

@article{codesim,
  title={CODESIM: Multi-Agent Code Generation and Problem Solving through Simulation-Driven Planning and Debugging},
  author={Islam, Md Ashraful and Ali, Mohammed Eunus and Parvez, Md Rizwan},
  journal={arXiv preprint arXiv:2502.05664},
  year={2025}
}

@article{shen2025optimizing,
  title={Optimizing LLM-Based Multi-Agent System with Textual Feedback: A Case Study on Software Development},
  author={Shen, Ming and Shu, Raphael and Pratik, Anurag and Gung, James and Ge, Yubin and Sunkara, Monica and Zhang, Yi},
  journal={arXiv preprint arXiv:2505.16086},
  year={2025}
}

@article{evomac,
  title={Self-evolving multi-agent collaboration networks for software development},
  author={Hu, Yue and Cai, Yuzhu and Du, Yaxin and Zhu, Xinyu and Liu, Xiangrui and Yu, Zijie and Hou, Yuchen and Tang, Shuo and Chen, Siheng},
  journal={arXiv preprint arXiv:2410.16946},
  year={2024}
}

@inproceedings{sreedhar2025simulating,
  title={Simulating cooperative prosocial behavior with multi-agent LLMs: Evidence and mechanisms for AI agents to inform policy decisions},
  author={Sreedhar, Karthik and Cai, Alice and Ma, Jenny and Nickerson, Jeffrey V and Chilton, Lydia B},
  booktitle={Proceedings of the 30th International Conference on Intelligent User Interfaces},
  pages={1272--1286},
  year={2025}
}

@article{wang2025talk,
  title={Talk structurally, act hierarchically: A collaborative framework for llm multi-agent systems},
  author={Wang, Zhao and Moriyama, Sota and Wang, Wei-Yao and Gangopadhyay, Briti and Takamatsu, Shingo},
  journal={arXiv preprint arXiv:2502.11098},
  year={2025}
}

@article{ted,
  title={The Truth Becomes Clearer Through Debate! Multi-Agent Systems with Large Language Models Unmask Fake News},
  author={Liu, Yuhan and Liu, Yuxuan and Zhang, Xiaoqing and Chen, Xiuying and Yan, Rui},
  journal={arXiv preprint arXiv:2505.08532},
  year={2025}
}

@inproceedings{thematiclM,
  title={Thematic-LM: A LLM-based Multi-agent System for Large-scale Thematic Analysis},
  author={Qiao, Tingrui and Walker, Caroline and Cunningham, Chris and Koh, Yun Sing},
  booktitle={Proceedings of the ACM on Web Conference 2025},
  pages={649--658},
  year={2025}
}

@article{multiagentbench,
  title={MultiAgentBench: Evaluating the Collaboration and Competition of LLM agents},
  author={Zhu, Kunlun and Du, Hongyi and Hong, Zhaochen and Yang, Xiaocheng and Guo, Shuyi and Wang, Zhe and Wang, Zhenhailong and Qian, Cheng and Tang, Xiangru and Ji, Heng and others},
  journal={arXiv preprint arXiv:2503.01935},
  year={2025}
}

@article{realworldbench,
  title={REALM-Bench: A Real-World Planning Benchmark for LLMs and Multi-Agent Systems},
  author={Geng, Longling and Chang, Edward Y},
  journal={arXiv preprint arXiv:2502.18836},
  year={2025}
}

@article{redteaming,
  title={Red-Teaming LLM Multi-Agent Systems via Communication Attacks},
  author={He, Pengfei and Lin, Yupin and Dong, Shen and Xu, Han and Xing, Yue and Liu, Hui},
  journal={arXiv preprint arXiv:2502.14847},
  year={2025}
}

@article{corba,
  title={CORBA: Contagious Recursive Blocking Attacks on Multi-Agent Systems Based on Large Language Models},
  author={Zhou, Zhenhong and Li, Zherui and Zhang, Jie and Zhang, Yuanhe and Wang, Kun and Liu, Yang and Guo, Qing},
  journal={arXiv preprint arXiv:2502.14529},
  year={2025}
}

@inproceedings{defense1,
  title={Improving alignment and robustness with circuit breakers},
  author={Zou, Andy and Phan, Long and Wang, Justin and Duenas, Derek and Lin, Maxwell and Andriushchenko, Maksym and Kolter, J Zico and Fredrikson, Matt and Hendrycks, Dan},
  booktitle={The Thirty-eighth Annual Conference on Neural Information Processing Systems},
  year={2024}
}

@article{defense2,
  title={Baseline defenses for adversarial attacks against aligned language models},
  author={Jain, Neel and Schwarzschild, Avi and Wen, Yuxin and Somepalli, Gowthami and Kirchenbauer, John and Chiang, Ping-yeh and Goldblum, Micah and Saha, Aniruddha and Geiping, Jonas and Goldstein, Tom},
  journal={arXiv preprint arXiv:2309.00614},
  year={2023}
}

@article{protocolsurvey,
  title={A survey of agent interoperability protocols: Model context protocol (mcp), agent communication protocol (acp), agent-to-agent protocol (a2a), and agent network protocol (anp)},
  author={Ehtesham, Abul and Singh, Aditi and Gupta, Gaurav Kumar and Kumar, Saket},
  journal={arXiv preprint arXiv:2505.02279},
  year={2025}
}

\end{document}